\documentclass[conference,compsoc]{IEEEtran}

\usepackage{relsize}  %

\usepackage{makecell} %

\usepackage{amssymb, amsmath}
\usepackage{graphicx} 
\usepackage{tabularx, threeparttable, booktabs}

\usepackage[disable]{todonotes} %
\usepackage{hyperref}
\usepackage{cleveref}

\usepackage{enumitem}
\setlist[description]{
    itemsep=0pt, %
    parsep=0pt, %
    leftmargin=0pt %
}
\setlist[itemize]{
    itemsep=2pt, %
    parsep=2pt, %
    leftmargin=10pt %
}
\setlist[enumerate]{
    itemsep=2pt, %
    parsep=2pt, %
}

\usepackage{tcolorbox}
\tcbuselibrary{skins}

\title{
Prompt Pirates Need a Map: Stealing Seeds helps Stealing Prompts 

}
\author{%
    \IEEEauthorblockN{%
        Felix~Mächtle\IEEEauthorrefmark{1},
        Ashwath~Shetty\IEEEauthorrefmark{2},
        Jonas~Sander\IEEEauthorrefmark{1},
        Nils~Loose\IEEEauthorrefmark{1},
        Sören~Pirk\IEEEauthorrefmark{2},
        Thomas~Eisenbarth\IEEEauthorrefmark{1}
    }\\[0.2em]
    $\vspace{-1em}$\IEEEauthorblockA{\IEEEauthorrefmark{1}Institute for IT Security, University of Lübeck, Germany\\
    \texttt{\{f.maechtle,\,j.sander,\,n.loose,\,thomas.eisenbarth\}@uni-luebeck.de}}\\
    $\vspace{-1em}$\IEEEauthorblockA{\IEEEauthorrefmark{2}Visual Computing and Artificial Intelligence, Kiel University, Germany\\
    \texttt{shettyashwath2010@gmail.com, sp@informatik.uni-kiel.de}}
}
\date{}

\newcommand{\ourapproach}{PromptPirate}

\newcommand{\bheading}[1]{{{\textbf{#1.}}}}

\begin{document}

\maketitle

\begin{abstract}
Diffusion models have significantly advanced text-to-image generation, enabling the creation of highly realistic images conditioned on textual prompts and seeds. Given the considerable intellectual and economic value embedded in such prompts, prompt theft poses a critical security and privacy concern. In this paper, we investigate prompt-stealing attacks targeting diffusion models. We reveal that numerical optimization-based prompt recovery methods are fundamentally limited as they do not account for the initial random noise used during image generation.
We identify and exploit a noise-generation vulnerability (CWE-339), prevalent in major image-generation frameworks, originating from PyTorch's restriction of seed values to a range of $2^{32}$ when generating the initial random noise on CPUs. Through a large-scale empirical analysis conducted on images  shared via the popular platform CivitAI, we demonstrate that approximately 95\% of these images' seed values can be effectively brute-forced in 140 minutes per seed using our seed-recovery tool, SeedSnitch.
Leveraging the recovered seed, we propose PromptPirate, a genetic algorithm-based optimization method explicitly designed for prompt stealing. PromptPirate surpasses state-of-the-art methods, i.e., PromptStealer, P2HP, and CLIP-Interrogator, achieving an 8–11\% improvement in LPIPS similarity. 
Furthermore, we introduce straightforward and effective countermeasures that render seed stealing, and thus optimization-based prompt stealing, ineffective.
We have disclosed our findings responsibly and initiated coordinated mitigation efforts with the developers to address this critical vulnerability.

\end{abstract}

\section{Introduction}

Over the past decade, machine learning techniques have been widely used in various areas of computer science~\cite{DBLP:conf/nips/VaswaniSPUJGKP17,loose2023madvex,machtle2025ocean,machtle2025trace}. Among these, diffusion models have emerged as a revolutionary approach to image generation~\cite{DBLP:conf/nips/HoJA20/StableDiffusion,song2020denoising,DBLP:journals/corr/ldm}.
These models have rapidly gained prominence due to their ability to generate highly realistic, cost-effective, and high-quality images conditioned on an input seed and a textual prompt.
Artists, designers and filmmakers have adopted them to craft compelling visuals, generate intricate animations, and build immersive virtual experiences. Notable examples include the award-winning, AI-generated artwork \textit{Théâtre d’Opéra Spatial} by Jason M. Allen~\cite{theatre_dopera_spatial}; the AI-generated short film \textit{The Crow} by Glenn Marshall, winning prizes at Cannes and Linz~\cite{heise_artificial_imagination}; and influential virtual personalities, such as the artificial influencer \textit{Lil Miquela}, who collaborates with global brands~\cite{lilmiquela_instagram}.

A central aspect of diffusion models is the textual prompt, which is an invaluable asset that defines the quality, specificity, and ultimately, the commercial potential of the generated content. The prompt provides critical semantic context, stylistic precision, and nuanced direction, guiding the model to produce compelling and commercially viable outputs. Mastery of prompt engineering has therefore become its own skill in fields such as advertising, entertainment and social media.
The growing economic significance of the expertise of generating the correct modifiers is evidenced by platforms like PromptBase~\cite{promptbase},  where specialized prompts are actively traded. Shen \emph{et al.}~\cite{DBLP:conf/uss/ShenQ0024/UsenixImagePromptStealing} estimated that the top 50 sellers on PromptBase collectively sold approximately 45,000 prompts over a nine-month period in 2022, generating roughly 186,525~USD in total revenue.

\begin{table}[t]
    \centering
    \begin{threeparttable}
        \caption{Seed knowledge is essential for  online prompt stealing. }
        \label{tab:intro:importance-of-seeds}
        \begin{tabular}{l|ccccc} \toprule

&\makecell{Original prompt \\ with seed $s$ \\ $\,$} &
\makecell{Stolen prompt \\ stolen seed $s$ \\ (\ourapproach{})} &
\makecell{No-seed \\attack \\ (P2HP~\cite{DBLP:conf/cvpr/MahajanRYS24/PromptingHardOrHardly})} &
 \\ \midrule
Prompt & \textit{A dog ...} & \textit{A dog ...} & \textit{An alien ...} \\
        \raisebox{0.09\linewidth-0.5em}[0em][0em]{Image} &\includegraphics[width=0.18\linewidth]{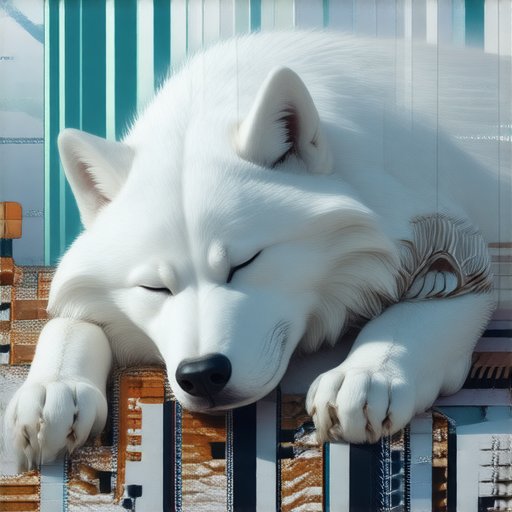}
        &    \includegraphics[width=0.18\linewidth]{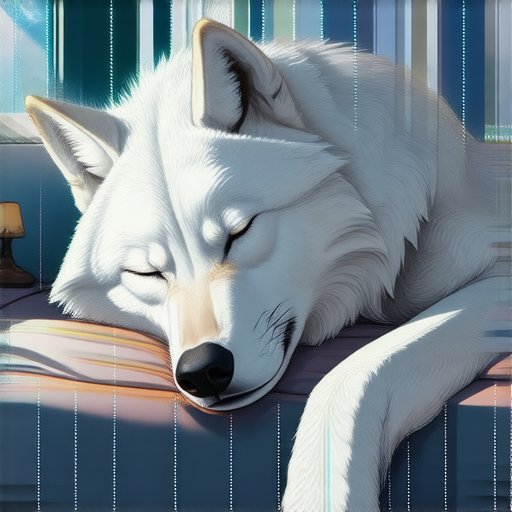} & 
        \includegraphics[width=0.18\linewidth]{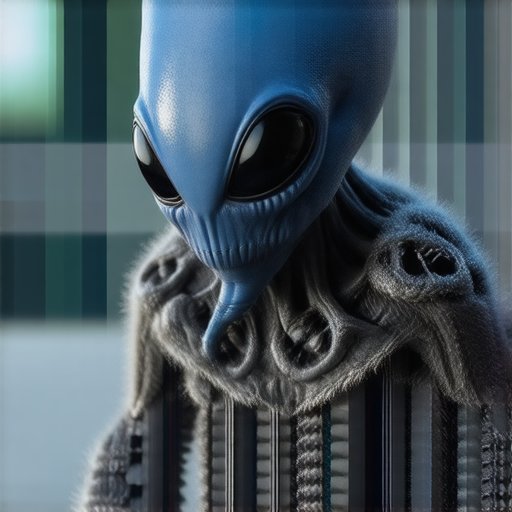} 

 \\ \bottomrule
        \end{tabular}
    \end{threeparttable}
\end{table}

Given the increasing economic and intellectual value associated with carefully crafted prompts, prompt-stealing represents a significant security challenge. If the unique style of a movie, artwork, or influencer's identity embedded within a prompt is compromised, it can easily be replicated and falsified, undermining authenticity and potentially leading to financial and reputational harm.  Consequently, recent research has increasingly focused on addressing and understanding the vulnerabilities associated with prompt-stealing. Recent work~\cite{DBLP:conf/uss/ShenQ0024/UsenixImagePromptStealing,DBLP:conf/cvpr/MahajanRYS24/PromptingHardOrHardly, DBLP:journals/cviu/CroitoruHIS24, clip_interrogator} has demonstrated that prompts can be stolen.

These works fall into two groups: \textit{Offline} and \textit{online} approaches.
To capture the style associated with an image, classification models have been proposed~\cite{DBLP:conf/uss/ShenQ0024/UsenixImagePromptStealing, DBLP:journals/cviu/CroitoruHIS24}. These models are expensively trained in an offline phase but during inference, these models, given an input image, rapidly generate a likelihood vector for a large set of predefined style modifiers, i.e., specific textual cues describing aesthetic attributes or visual styles such as "cinematic lighting," "oil painting," or "cyberpunk". 
However, the offline-phase of such classifiers is very resource-intensive. They require thousands of images rendered under different seed values and a wide variety of prompts that exhaustively combine relevant style modifiers. This combinatorial explosion makes comprehensive coverage practically infeasible. Furthermore, diffusion models are highly sensitive to architectural adjustments and changes to the training dataset. Any update to the underlying generation model invalidates prior training, necessitating full retraining from scratch. This inability to generalize across versions severely limits the utility and scalability of classifier-based approaches for prompt reversing tasks.
In comparison, online approaches~\cite{DBLP:conf/cvpr/MahajanRYS24/PromptingHardOrHardly, clip_interrogator} utilize the diffusion models directly to optimize a candidate prompt and eliminate the need for an offline phase. Thus, while these methods do not require an offline phase and generalize to different model types and versions, the online phase is longer.

However, current online techniques overlook the critical influence of the initial seed. %
Recent research by Xu \emph{et al.}~\cite{DBLP:conf/wacv/XuZS25/Golden_Seed} emphasizes the importance of seeds in image generation. %
Because the seed generates the initial noise pattern from which the image subsequently evolves, the seed inherently embeds unique characteristics into the final output. Therefore, images produced with different seeds (even when using the same prompt) can appear entirely distinct to human observers. 
All online methods use a distance function to measure the similarity between the current candidate and the target image, i.e., a loss value. %
However, these distance functions rely more on the seed than on the visual similarity because the seed transforms the image significantly. As shown in \Cref{tab:intro:importance-of-seeds}, the similarity metrics employed by online techniques do not work reliably without knowing the seed.

We demonstrate that online prompt stealing cannot be effectively performed without knowledge of the initial randomness, specifically the seeds used by the pseudorandom number generators (PRNGs). 
However, if the seed is chosen from a relatively small range of possibilities, such as $2^{32}$, we show that it can be practically recovered in $140$ minutes, using our proposed tool, SeedSnitch. Using a narrow seed range has a well-documented history in software projects~\cite{debianCVE2008, bitcoinAndroid2013}, frequently resulting in severe security impacts, known as CWE-339 ("Small Seed Space in PRNG"). We have identified such critical security vulnerabilities in multiple implementations of modern diffusion-based image generators. Specifically, these implementations either artificially limit the seed range (e.g., 0–100,000) or utilize the PyTorch PRNG on the CPU, whose implementation caps the seed at 32 bits. Thus, even if larger seeds are generated, only the lower 32 bits are used.
Consequently, only 32 bits need to be brute-forced, making both scenarios effectively susceptible to brute-force attacks.

Leveraging recovered seeds, we propose a novel online method explicitly designed to exploit knowledge of the initial randomness, thus enhancing prompt recovery accuracy. Our approach outperforms existing online and offline-based prompt-stealing methods in visual reconstruction. %
Consequently, this work offers a practical and efficient solution for seed extraction and subsequent prompt recovery, underscoring both the vulnerabilities present in current systems and the paramount importance of robust seed security practices.

To foster transparency, reproducibility, and facilitate security research, we open-source PromptPirate and SeedSnitch on GitHub\footnote{\url{https://github.com/UzL-ITS/Prompt-Pirate}}.

To disclose this issue responsibly, we engaged in the process of a responsible disclosure and obtaining CVEs. To the best of our knowledge, this work is the first to identify and exploit a CWE‑339 security vulnerability in an image generation model, marking an important milestone in both the study of Diffusion Models and prompt security.

To summarize, our contributions are:
\begin{itemize} %
    \item We present PromptPirate, a novel, effective and highly reliable online method for accurately stealing image prompts. Our comprehensive empirical evaluation shows that prompts extracted through PromptPirate produce image results significantly more similar to the image generated by the secret target prompt compared to prompts stolen via previous attacks.

    \item To tackle the complex non-differentiable problem of extracting the textual prompt modifiers from the output image of diffusion models, we propose the first optimization approach based on genetic algorithms as a key driver of PromptPirate, 
    carefully designed to accurately steal prompt modifiers. %

    \item Based on prior work by Xu \emph{et al.}~\cite{DBLP:conf/wacv/XuZS25/Golden_Seed} we systematically evaluate the influence of the initial noise seed, used by Diffusion Models, on the generated image. We found that standard loss functions (i.e., Latent-MSE, LPIPS, CLIP) used by previous online attacks are more sensitive to different seeds than to different prompt modifiers, making the extraction success strongly dependable on the initial noise seed. Consequently, we show that known online extraction attacks get highly unreliable and mainly produce poor prompt extractions when the initial noise seed is unknown.

    \item As part of PromptPirate, we propose SeedSnitch, a powerful, brute-force based seed stealing attack  that boosts the reliability and quality of the extracted prompts. While previous work manually limited the range of possible seeds to enable extraction, SeedSnitch successfully extracts seeds from unmodified and up-to-date mainstream implementations like Stable Diffusion 3.5.
    
    \item We present a large real-world case study demonstrating the efficiency, effectiveness, and reliability of SeedSnitch. Concretely, we found SeedSnitch successfully extracts 95\% of used seeds from realistic output images in 140 minutes per seed.
    
    \item As part of SeedSnitch, we identify and responsibly disclose a pervasive CWE‑339 security vulnerability in a wide range of popular image diffusion implementations.

    \item While mitigating the success of SeedSnitch and PromptPirate is not possible for already published images, we finally discuss countermeasures to better protect the initial noise seed and consequently the prompt used to generate public images.
    
\end{itemize}

\section{Preliminaries}

We introduce our notations in \Cref{tab:notation}.

\begin{table}[t]
\centering
\caption{Notation}
\label{tab:notation}
\begin{tabular}{ll}
\toprule
Symbol & Definition \\
\midrule
$\mathcal{N}(0, \mathbf{I})$ & Normal distribution, mean 0, identity covariance $\mathbf{I}$ \\
$\mathcal{M}_\theta$ & Conditional diffusion model \\
$\mathcal{G}$ & Complete image generation function \\
$z$ & Image in latent space \\
$I$ & Image in pixel space \\
$\mathcal{E}(I)$ & Encoder, transforming images to latent space \\
$\mathcal{D}(z)$ & Decoder, transforming latents to images \\
$s$ & Seed value \\
$\epsilon_s$ & Initial noise, generated by seed $s$ \\
\bottomrule
\end{tabular}
\end{table}

\subsection{Diffusion Models}

Diffusion models \cite{DBLP:conf/nips/HoJA20/StableDiffusion, song2020denoising} are generative models that create images through an iterative denoising process. The forward diffusion process gradually corrupts clean data $x_0$ by adding Gaussian noise over $T$ timesteps:
\begin{equation}
q(x_t | x_{t-1}) = \mathcal{N}(x_t; \sqrt{\alpha_t} x_{t-1}, (1-\alpha_t) \mathbf{I})
\end{equation}
where $\{\alpha_t\}_{t=1}^T$ is a predefined noise schedule that controls the amount of noise added at each step.

The reverse diffusion process learns to denoise by training a neural network $\epsilon_\theta$ to predict the noise at each timestep:
\begin{equation}
p_\theta(x_{t-1} | x_t, p) = \mathcal{N}\left(x_{t-1}; \mu_\theta(x_t, t, p), \sigma_t^2 \mathbf{I}\right)
\end{equation}
where $p$ represents conditioning information (e.g., text embeddings) and $\theta$ denotes the model parameters. During generation, the model starts from pure noise $x_T \sim \mathcal{N}(0, \mathbf{I})$ and iteratively applies the learned denoising steps to produce a clean image $x_0$. We denote the full generation process in pixel space as $x_0 = \mathcal{M}_\theta(x_T, p)$, where $\mathcal{M}_\theta$ represents the learned denoising model conditioned on $p$.

\subsection{Latent Diffusion Models}

Latent Diffusion Models (LDMs), such as Stable Diffusion, operate in a compressed latent space rather than directly on high-resolution pixel space for computational efficiency. An encoder $\mathcal{E}: \mathbb{R}^{H \times W \times 3} \rightarrow \mathbb{R}^{h \times w \times c}$ maps input images to a lower-dimensional latent representation, where typically $h, w \ll H, W$ (e.g., $64 \times 64$ latents for $512 \times 512$ images). The diffusion process is then applied to these latent representations $z = \mathcal{E}(I)$.

After the denoising process completes in latent space, a decoder $\mathcal{D}: \mathbb{R}^{h \times w \times c} \rightarrow \mathbb{R}^{H \times W \times 3}$ reconstructs the final image: $I = \mathcal{D}(z_0)$. This approach significantly reduces computational costs while maintaining high-quality generation. In this case, generation is performed in latent space as $z_0 = \mathcal{M}_\theta(z_T, p)$, followed by decoding the final image as $I = \mathcal{D}(z_0)$. %

\subsection{Generation Pipeline and Notation}

Let $p$ be the conditioning to the diffusion model (for our setting we consider it to be a text prompt), $s$ a random seed, $\mathcal{M}_\theta$ the conditional diffusion model with parameters $\theta$, and $\text{PRNG}(s) = \epsilon_s \sim \mathcal{N}(0, \mathbf{I})$ the initial noise tensor derived from seed $s$. The generation process proceeds as:
\[
\begin{aligned}
\epsilon_s &= \text{PRNG}(s) \\
z_T &= \epsilon_s \\
z_0 &= \mathcal{M}_\theta(z_T, p) \\
I &= \mathcal{D}(z_0)
\end{aligned}
\]
The complete generation function can be written as:
\[
I = \mathcal{G}(p, s, \theta) = \mathcal{D}(\mathcal{M}_\theta(\text{PRNG}(s), p))
\]

\subsection{Prompt Stealing in Diffusion Models}

Prompt stealing has emerged as a critical security and economic concern in the era of generative AI. 
Formally prompt stealing can be described in the following: Given a target image $I_{\text{target}} = \mathcal{G}(p_{\text{orig}}, s_{\text{orig}}, \theta)$ generated by a known model $\mathcal{G}$, the objective of prompt stealing is to recover a prompt $p^*$ that yields visually similar outputs:
\begin{equation}
p^* = \arg\min_{p \in \mathcal{P}} \mathcal{L}(\mathcal{G}(p, s', \theta), I_{\text{target}})
\end{equation}
where $\mathcal{P}$ is the space of textual prompts, $s'$ is a seed (fixed or varied), and $\mathcal{L}$ is a similarity loss, e.g., CLIP~\cite{DBLP:conf/icml/RadfordKHRGASAM21/CLIP}, LPIPS~\cite{DBLP:conf/cvpr/ZhangIESW18/LPIPS} or mean squared error (MSE).

\noindent\bheading{Offline Approaches}
These methods generate a prompt directly using pre-trained models, thus they require a long offline phase, where they generate thousands of images using different prompts and seeds. However, therefore during inference they avoid iterative optimization. A prominent  example is PromptStealer by Shen \emph{at al.}~\cite{DBLP:conf/uss/ShenQ0024/UsenixImagePromptStealing}, which uses a generative model to generate the subject part of the prompt. To predict style modifiers, PromptStealer uses a classification-based model that was trained on a large set of images that were annotated with various modifiers. Thus, both the subject and the modifiers are predicted directly. %

\noindent\bheading{Online Approaches}
In contrast, online methods attempt to incrementally refine an initial prompt until it matches the target image. 
One such method is CLIP Interrogator~\cite{clip_interrogator}. It operates by initially generating a subject description using a large vision language model \cite{DBLP:conf/icml/0001LXH22/Blip}. It then utilizes CLIP \cite{DBLP:conf/icml/RadfordKHRGASAM21/CLIP} to embed the target image and compares this to the precomputed embeddings of candidate textual modifiers. This is possible because CLIP is trained to project text and images into the same embedding space. Each candidate modifier is ranked according to its similarity score, meaning that modifiers with higher similarity scores are considered more relevant and descriptive for constructing the final prompt. The final prompt is constructed by greedily combining the subject description with top-ranked modifiers, selecting only those combinations that incrementally improve the overall similarity to the image embedding. %

Recent work,   P2HP~\cite{DBLP:conf/cvpr/MahajanRYS24/PromptingHardOrHardly}, operates directly within the diffusion model's continuous text-embedding input space to refine prompt representations. They iteratively optimize a latent token embedding to minimize the distance between the generated and target images. It uses the L-BFGS~\cite{shanno1970conditioning} optimizer, which optimizes discrete token embeddings and ensures that the optimized tokens remain within the space of valid words. All generated prompts are stored during this iterative optimization, and subsequently, the best prompt is selected based on CLIP similarity. However, both approaches don't consider the importance of the seed in the prompt stealing process, hence they often gets stuck in local minima, hampering the results.

\section{\ourapproach{}}

\begin{figure}
    \centering
    \includegraphics[width=\linewidth]{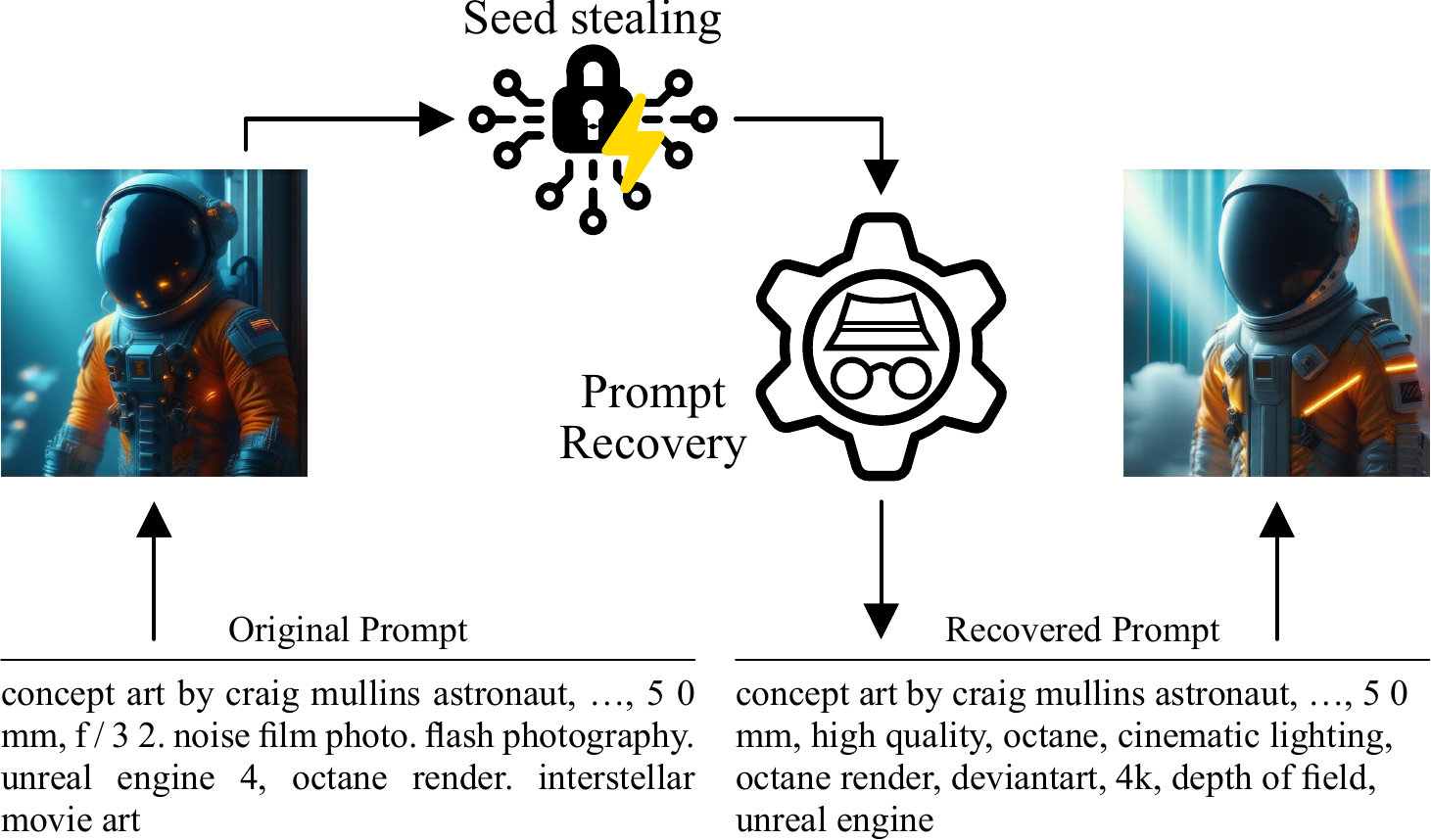}
    \caption{Our pipeline for prompt recovery.}
    \label{fig:our-pipeline}
\end{figure}

This section outlines our threat model, vulnerabilities found in stable diffusion implementations, research methodology and our proposed approach to seed recovery. The pipeline of \ourapproach{} is shown in \Cref{fig:our-pipeline}.

\subsection{Threat Model}
In our threat model and similar to previous work, we assume an attacker that wants to steal a secret prompt which was used together with an open available generative diffusion model to generate a public image. 
The attacker specifically seeks to recover a prompt that reproduces the visual outcome (e.g., style, composition, salient attributes) to generate new images visually indistinguishable or closely resembling the original.
Such a setting naturally appears in common real-world scenarios like prompt marketplaces and AI-generated art, where the underlying style holds the intellectual and monetary value.

We assume the attacker knows the exact generative diffusion model including its encoder as well as the hyperparameters and the noise generation strategy used to generate the public image. Previous work \cite{DBLP:journals/corr/abs-2503-10718/DetectionOfAIModel} showed that the model can also be inferred or validated from a generated image using classification methods, and the relevant hyperparameters and scheduler choices are often derivable from foundational publications and repositories. This assumption is practical in today’s deployment landscape, where a small number of models dominate usage (e.g., on CivitAI in August 2025, only seven distinct models appear among the 50 most-liked uploads out of roughly a few dozen models listed), allowing an attacker to narrow the search to a small candidate set. Based on typical real-world scenarios, we assume the seed for the initial noise generation to be private.

\subsection{Implementation Vulnerabilities}
\label{sec:rng-vulnerabilities}

\begin{table}[]
\setlength{\tabcolsep}{1.6pt}
    \centering

\begin{threeparttable}
    \caption{Identified Vulnerabilities in Popular Image Generation Tools}
    \label{tab:vulnerable-products}
    
    \begin{tabularx}{\linewidth}{lXrr} \toprule
       Name  & Description & Stars & Vuln.  \\ \midrule
       AUTOMATIC1111\cite{automatic1111_vuln}  & Popular web UI for Stable Diffusion & 153K & V2 \\ 
       ComfyUI~\cite{comfyui_vuln}  & Workflow interface for image synthesis & 78.1K & V2  \\
       Diffusers~\cite{huggingface_diffusers} & Hugging Face’s widely adopted library for diffusion-based generation & 30.7K & V2 \\
       Easy Diffusion~\cite{easydiffusion_vuln}  & Beginner-friendly image generation interface & 9.9K & V1 \\
       InvokeAI~\cite{invokeai_vuln}  & Powerful toolkit for professionals& 25.2K & V2  \\
       \makecell[l]{Stable  \\Diffusion~3.5~\cite{stabilityai_sd3_infer}}  &  Reference implementation & 1.2K  & V1/V2 \\ 
       \bottomrule
    \end{tabularx}
    
\end{threeparttable}
\end{table}

In our analysis of various PyTorch-based image generation products, we identified two types of vulnerabilities related to the generation of initial random noise, critically affecting image generation security. These vulnerabilities were found in five different, widely used, and highly starred GitHub projects, underscoring the breadth and severity of the issue across the ecosystem. 

\noindent\bheading{Limited Seed Range (V1)} The first common vulnerability observed is an overly restrictive seed range. In two analyzed implementations, the seed used to initialize the random number generator is chosen from a severely limited numerical interval, exemplified by patterns such as:
\begin{verbatim}
seed_num = torch.randint(0, 100000)
\end{verbatim}
This significantly restricted range severely limits the entropy required for secure random number generation, drastically reducing the search space available to an attacker (e.g., $2^{16.6}$ for a range of 0--100{,}000, as implemented in the Stable Diffusion 3.5 reference implementation.

\noindent\bheading{Weak CPU-based RNG Algorithm (V2)} Even if the limited seed range is expanded, a second prevalent issue arises from the explicit or implicit use of PyTorch’s CPU-based random number generation:
\begin{verbatim}
torch.randn(..., device="cpu")
\end{verbatim}
This implicitly selects PyTorch's default MT19937 algorithm, which utilizes only the lower 32 bits of the provided seed\footnote{\url{https://github.com/pytorch/pytorch/blob/d68d4d31f4824f1d1e0d1d6899e9879ad19b0754/aten/src/ATen/core/MT19937RNGEngine.h\#L158}}. Consequently, as documented within PyTorch's internal implementation, seeds exceeding 32 bits undergo truncation:
\begin{verbatim}
data_.state_[0] = seed & 0xffffffff;
\end{verbatim}
Thus, even seemingly extensive seed spaces effectively reduce to at most $2^{32}$ unique seeds, causing distinct 64-bit seeds to produce identical random number sequences, for any integers $s,\,\alpha \in \left[0, 2^{32}\right]$:
\[
\texttt{MT19937}(s) = \texttt{MT19937}(s + \alpha * 2^{32})
\]
This fundamental limitation greatly reduces the security benefits of larger seed spaces, leaving these implementations vulnerable to brute-force seed recovery attacks.

\Cref{tab:vulnerable-products} categorizes these vulnerabilities and summarizes the affected implementations. Notably, only the macOS version of AUTOMATIC1111 is vulnerable and the Hugging Face Diffusers library only in certain configurations.
The presence of these vulnerabilities across multiple products directly enables practical brute-force attacks, significantly compromising the confidentiality and integrity of prompt-based image generation. %

We followed CERT/CC’s 45-day coordinated disclosure process, and the 45-day disclosure window has now elapsed. For each affected implementation, we prepared a technical summary and a working proof of concept and contacted core maintainers via the security or personal email addresses listed on their GitHub profiles and, where available, via Slack. Where possible, we also filed reports through \url{huntr.com}. Hugging Face has acknowledged our report regarding Diffusers and indicated that, while no immediate change is planned, mitigations may be incorporated in a future release. Easy Diffusion has engaged in discussion, and we have provided additional technical details to support their assessment. The remaining projects have not yet responded.

\subsection{SeedSnitch}
\label{sec:seed-recovery-formal}

The seed is a fundamental component of the diffusion-based image generation process, directly influencing the initial random noise from which an image is subsequently derived. 
Formally, the seed \(s\) determines the initial latent noise vector \(\epsilon_s\), which serves as the starting point $z_T$ for the iterative denoising steps of the diffusion model. Crucially, the initial noise state significantly impacts the structure of the resulting image, embedding a unique signature within the image’s latent representation.

Recently Xu \emph{et al.}~\cite{DBLP:conf/wacv/XuZS25/Golden_Seed}, proposed training a classifier to predict seed values within a fixed range (i.e., 1024 possible seeds) and achieved near-perfect accuracy. Their findings suggest that generated images possess unique and identifiable fingerprints originating from their seeds.

Based on these insights, we present SeedSnitch, a new, lightweight attack that recovers the seed used in image generation. Unlike previous approaches that rely on classifier training, SeedSnitch directly exploits the latent structure of the generated image to recover the seed in a fully deterministic, training-free manner. We hypothesize that a robust and measurable correspondence exists between a seed and its resulting latent image representation.
Thus, we propose a simpler, yet effective, approach to seed recovery based on latent vector comparisons.
Given a target image \(I\) (with latent representation \(z_0\) inferred using the diffusion models encoder $\mathcal{E}(I) = z_0$) and a candidate seed \(s\), we compute the latent representation \(\epsilon_s = PRNG(s)\) derived from the  candidate seed.
The similarity between the target latent and the candidate seed latent is quantified using the Mean Squared Error (MSE) loss defined as:
\[
\text{MSE}(z_0, \epsilon_s) = \frac{1}{n}\sum_{i=1}^{n}(z_{0, i} - \epsilon_{s, i})^2
\]
where \(n\) denotes the dimensionality of the latent representation, and \(z_{0, i}\) and \(\epsilon_{s, i}\) represent the \(i\)-th element of the latent vectors, respectively.

\subsection{Prompt Modifier Recovery via Genetic Optimization}

Accurately reconstructing stylistic prompt modifiers presents a unique challenge.
Due to the high non-linearity of diffusion models, traditional gradient-based optimization methods are ineffective at recovering discrete tokens. This ineffectiveness arises because, although a model can easily be guided to generate a specific output image, the latent representation that produces this image often results from a local minimum. Hence, the corresponding embedding does not correspond directly to discrete text token embedding. Consequently, mapping back from image space to textual space requires projecting onto the nearest token embedding, which distorts or completely loses the original visual characteristics of the generated image. Existing research~\cite{DBLP:conf/cvpr/MahajanRYS24/PromptingHardOrHardly, DBLP:conf/nips/WenJKGGG23/PEZ} attempts to address this issue by using algorithms that continuously map latent vectors back to the nearest available token embeddings.

To overcome this challenge, we propose an optimization approach based on Genetic algorithms (GAs) and specifically designed for prompt modifier stealing. GAs are known for their efficacy in solving complex, non-differentiable problems~\cite{felix2025autostub} and offer an ideal framework for systematically exploring the latent space. %

Formally, our goal is to reconstruct the optimal combination of stylistic modifiers given a known prefix (e.g., main subject) and a recovered seed. We define the problem as minimizing the latent space discrepancy between the original target image and an image generated from candidate modifiers:
\begin{equation}
\text{argmin}_{m \in \mathcal{SM}} \quad \text{MSE}(z_{0}, z_{0,m,s})
\end{equation}

where $m$ represents a candidate modifier set, $\mathcal{SM}$ denotes the space of possible stylistic  modifier combinations, $z_0$ is the latent representation of the original image, and $z_{0,m,s}$ is the latent representation generated using modifier set $m$ and the recovered seed $s$.

Our genetic algorithm is structured as follows:

    \noindent\textbf{1) Initialization:} We generate an initial population of candidate modifier sets. Each candidate consists of between $3$ and $12$ randomly selected modifiers. To reduce noise and accelerate convergence, we restrict candidate modifiers to those appearing in at least $1\%$ of training prompts in the dataset of Shen \emph{et al.}~\cite{DBLP:conf/uss/ShenQ0024/UsenixImagePromptStealing}.

    \noindent\textbf{2) Fitness Evaluation:} The fitness of each candidate modifier set is evaluated by generating an image using the candidate prompt (composed of the known prefix and the candidate modifiers, i.e., \emph{subject, modfier$_1$, modifier$_2$, .., modifier$_n$})
    and the previously recovered seed. To ensure efficient image generation, we use the Stable Diffusion 3.5 Large Turbo model, which requires only 1/10 the number of denoising iterations compared to its standard counterparts.
     We measure fitness using the Mean Squared Error (MSE) loss between the latent representations of the generated candidate image and the target image:
    \[
    \text{Fitness}(m) = \text{MSE}(z_{0}, z_{0,m,s})
    \]
    where \(z_{0}\) is the latent representation of the original image, and \(z_{0,m,s}\) is the latent representation generated using modifier set \(m\) with the recovered seed \(s\).

    \noindent\textbf{3) Selection:} We use tournament selection~\cite{DBLP:conf/foga/GoldbergD90}, where subsets of candidates compete and the best is selected,
    to probabilistically choose high-performing candidates for reproduction. This encourages the propagation of beneficial traits within the population.

    \noindent\textbf{4) Crossover:} A variable-length one-point crossover operation recombines modifiers from two random parent candidates by independently selecting a random cut-point within each parent and then concatenating the segments before the cut from the first parent with the segments after the cut from the second parent. The operation does not cut within a single modifier.
    For example, given two parents \( [a_1, a_2 \mid a_3, a_4] \) and \( [b_1 \mid b_2, b_3] \), where the vertical bar indicates the chosen crossover points, the resulting offspring would be \( [a_1, a_2, b_2, b_3] \).
    If the offspring length violates predefined length constraints, it is adjusted by truncating excess tokens or padding with additional modifiers.

    \noindent\textbf{5) Mutation:} Mutation introduces variability within the population by probabilistically replacing, inserting, or deleting individual random modifiers within a candidate. The probabilities of these mutation operations are set to $0.15$ for replacement, $0.03$ for insertion, and $0.02$ for deletion.

    \noindent\textbf{6) Elitism:} In each generation, the top $5\%$ of candidates (the \emph{elite}) with the best fitness scores are copied unchanged into the next generation. This ensures that the highest-performing solutions discovered so far are preserved and not lost due to genetic operations such as mutation or crossover. %

This iterative process is repeated for $25$ generations, each consisting of $150$ individuals. At each generation, the best-performing candidate (lowest MSE loss) is tracked, resulting in the final optimal modifier set that best reconstructs the stylistic elements of the original prompt.

\section{Evaluation}

In this section, we present an empirical evaluation of our proposed techniques and vulnerabilities identified. Our evaluation focuses on understanding the impact of seed and modifier variations on generated images, assessing the effectiveness of our seed recovery method, and comparing our approach against state-of-the-art prompt-stealing methods. Additionally, we provide a practical case study based on real-world data in \Cref{sec:case-study-civitai}.

\subsection{Impact of Seed vs. Stylistic Modifiers on Image Similarity}
\label{sec:eval:seed-vs-modifier}

\begin{table}[t]
    \centering
    \begin{threeparttable}
    \caption{%
    Example images illustrating the impact of varying the seed (DSSM) versus a modifier (SSDM) on loss metrics (lower values indicate higher similarity). DSSM consistently produces greater divergence. %
    }
    \label{tab:example-images-ssdm-dssm}
        \begin{tabular}{c|cc} \toprule
        Original Image & SSDM & DSSM \\ \midrule
        \includegraphics[width=0.15\linewidth]{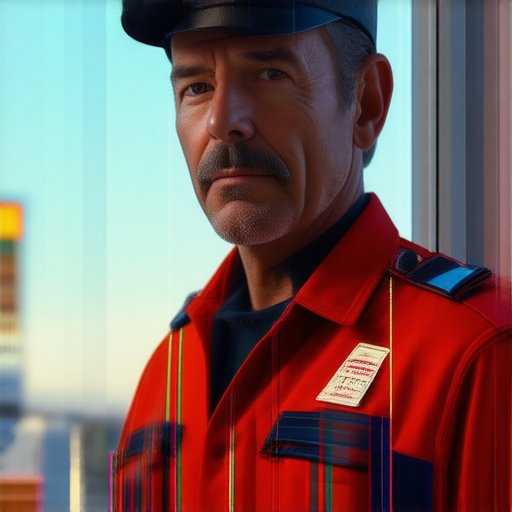} 
        & 
        \includegraphics[width=0.15\linewidth]{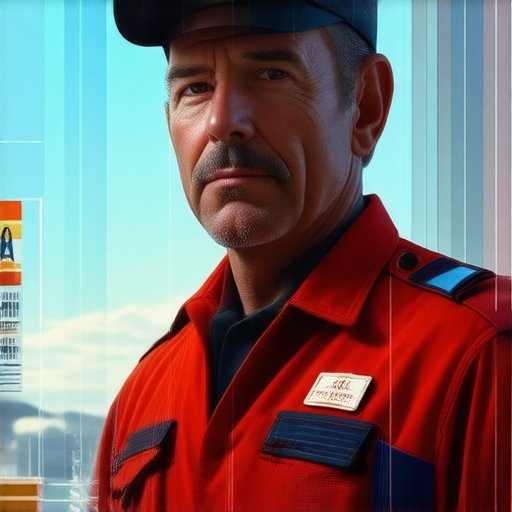} 
        \raisebox{0.075\linewidth-0.5em}[0em][0em]{\makecell[r]{MSE: 0.02 \\ LPIPS: 0.28 \\ CLIP: 0.22}} &
        \includegraphics[width=0.15\linewidth]{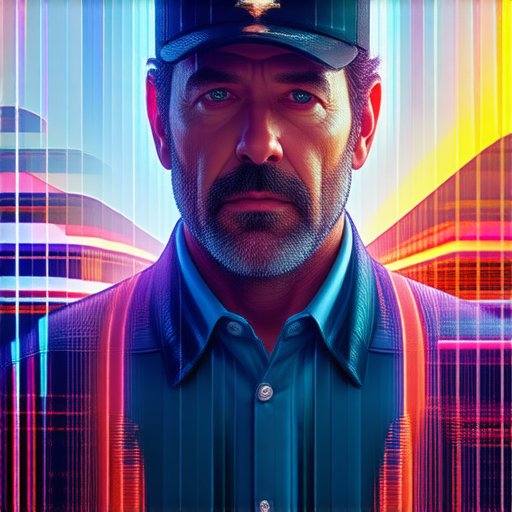} 
        \raisebox{0.075\linewidth-0.5em}[0em][0em]{\makecell[r]{MSE: 0.15 \\ LPIPS: 0.78 \\ CLIP: 0.28}} \\

        \includegraphics[width=0.15\linewidth]{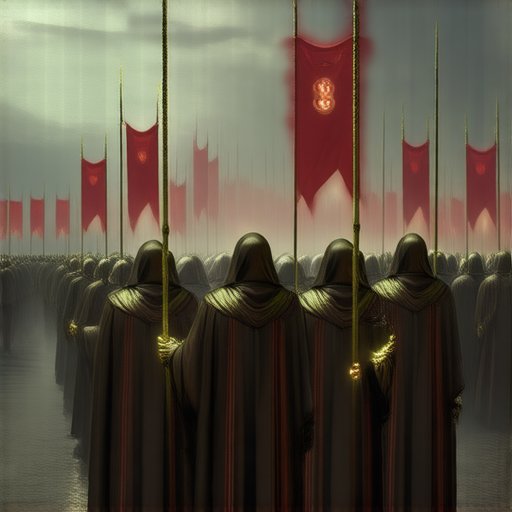} & 
        \includegraphics[width=0.15\linewidth]{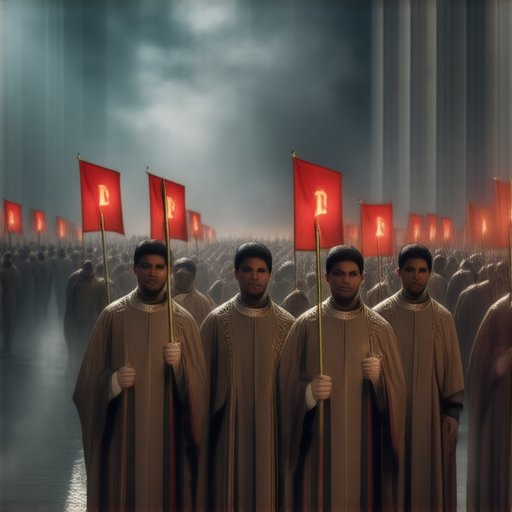} 
        \raisebox{0.075\linewidth-0.5em}[0em][0em]{\makecell[r]{MSE: 0.03 \\ LPIPS: 0.54 \\ CLIP: 0.22}} &
        \includegraphics[width=0.15\linewidth]{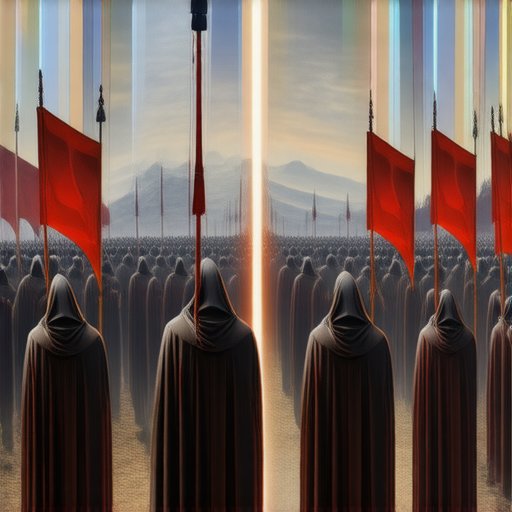} 
        \raisebox{0.075\linewidth-0.5em}[0em][0em]{\makecell[r]{MSE: 0.06 \\ LPIPS: 0.70 \\ CLIP: 0.12}} \\

        \includegraphics[width=0.15\linewidth]{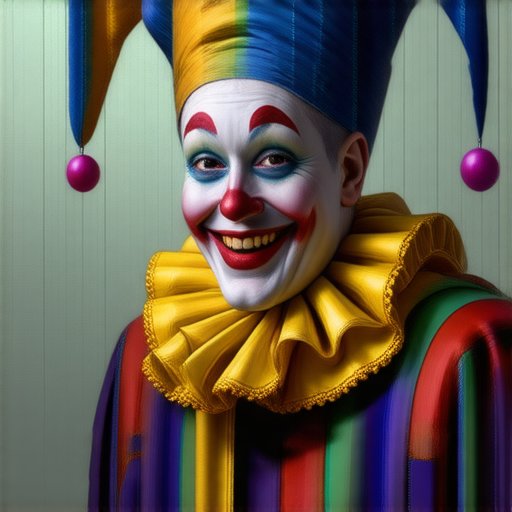} & 
        \includegraphics[width=0.15\linewidth]{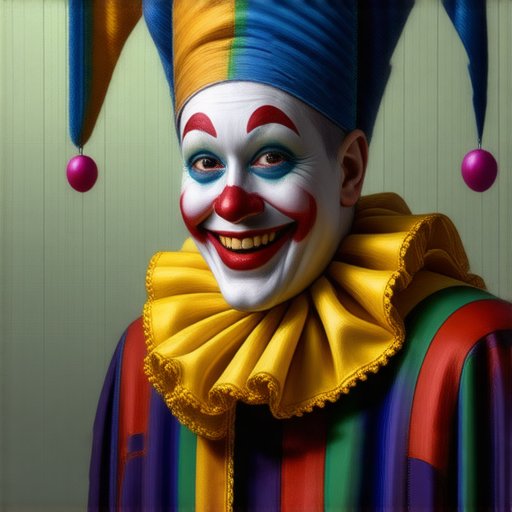} 
        \raisebox{0.075\linewidth-0.5em}[0em][0em]{\makecell[r]{MSE: 0.01 \\ LPIPS: 0.14 \\ CLIP: 0.01}} &
        \includegraphics[width=0.15\linewidth]{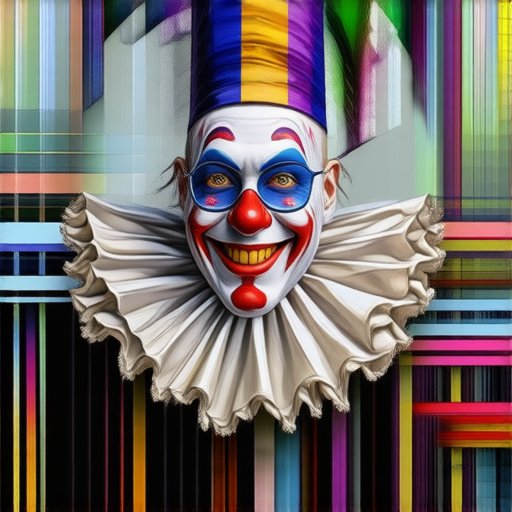} 
        \raisebox{0.075\linewidth-0.5em}[0em][0em]{\makecell[r]{MSE: 0.13 \\ LPIPS: 0.76 \\ CLIP: 0.07}} \\

\bottomrule
        
    \end{tabular}
    \end{threeparttable}
    
\end{table}

First, we will analyze the impact that different seeds have on the generated images. As shown in \Cref{tab:example-images-ssdm-dssm}, the output images vary greatly depending on the chosen seeds. Therefore, we hypothesize that the seed may have a greater impact on the final image than a single modifier would.

\noindent\bheading{Experiment Design}
To investigate this, we implemented the following experimental setup:
We selected $N=1,000$ prompts from the Shen \textit{et al.}~\cite{DBLP:conf/uss/ShenQ0024/UsenixImagePromptStealing} dataset. For every prompt, we generated two sets of image pairs: \textit{Same Seed, Different Modifiers} (SSDM), where we varied a single stylistic modifier while keeping the seed constant, and \textit{Different Seed, Same Modifiers} (DSSM), where we altered the seed but kept modifiers unchanged. Those images were compared to the original target image with correct seed and correct modifiers. This procedure produced 1,000 paired image comparisons, with each prompt yielding one pair. Each pair included a comparison of the original image with both the altered seed image and the altered modifier image.

To quantify perceptual and semantic differences between generated image pairs, we employed three widely used metrics:
\begin{itemize}
\item \textbf{CLIP Distance (CLIP)} measures semantic similarity between images using embeddings generated by  OpenAI's Contrastive Language-Image Pretraining (CLIP) model~\cite{DBLP:conf/icml/RadfordKHRGASAM21/CLIP}. CLIP itself is a neural network trained on diverse image-text pairs to jointly embed visual and textual information into a shared latent space, enabling cross-modal similarity assessment. We use it to embed both generated images and compare the cosine similarity between the two image embeddings, which reflects semantic closeness; lower values indicate higher semantic similarity.
\item \textbf{Learned Perceptual Image Patch Similarity (LPIPS)} quantifies perceptual differences by leveraging deep neural network activations~\cite{DBLP:conf/cvpr/ZhangIESW18/LPIPS}. LPIPS evaluates the perceptual similarity as humans perceive, with lower values indicating visually similar images.
\item \textbf{Latent Mean Squared Error (Latent-MSE)} computes the average squared differences between images in the latent representation space of the generative model. Lower values indicate greater similarity at the latent space level.
\end{itemize}

For a pair of images $I^{ref}$ and $I^{mod}$, the metrics are defined as:
\[
\small
\begin{aligned}
\text{CLIP}(I^{ref}, I^{mod})&=1-\cos\!\bigl(f_{\text{CLIP}}(I^{ref}),f_{\text{CLIP}}(I^{\text{mod}})\bigr),\\
\text{LPIPS}(I^{ref}, I^{mod})&=f_\text{LPIPS}(I^{ref}, I^{mod}),\\
\text{Latent-MSE}(z_0^{, ref}, z_0^{, mod})&=\frac1n\sum_{i=1}^n \bigl(z_{0, i}^{ref} - z_{0, i}^{mod}\bigr)^2,
\end{aligned}
\]
 $f_{\text{CLIP}}(\cdot)$ denotes CLIP embedding extraction, $f_\text{LPIPS}(I^{ref}, I^{mod})$ is the perceptual loss \cite{DBLP:conf/cvpr/ZhangIESW18/LPIPS} applied to the unchanged reference and modified image pairs and $z_0^{ref} = 
\mathcal{E}(I^{ref}), z_0^{\text{mod}}= 
\mathcal{E}(I^{mod})$ correspond to latent vectors of the compared image pairs.

\begin{figure}[t]
    \centering
    \includegraphics[width=0.3\linewidth]{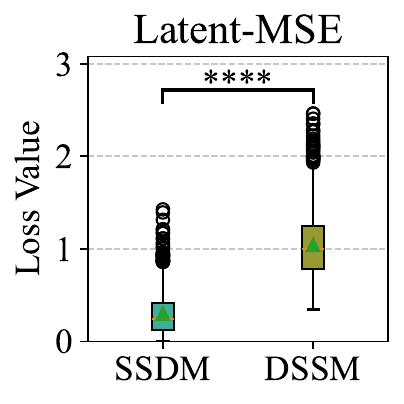}
    \includegraphics[width=0.3\linewidth]{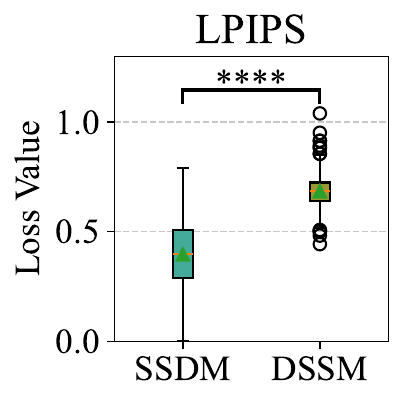}
    \includegraphics[width=0.3\linewidth]{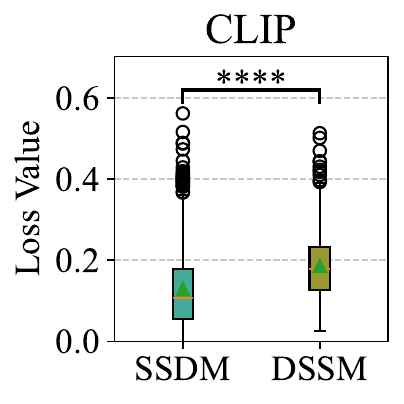}
    
    \caption{Comparison of perceptual, structural, and semantic differences between images generated with the Same Seed but Different Modifiers (SSDM) and Different Seed but Same Modifiers (DSSM). Lower values correspond to more similar images. The boxplots illustrate that variations in the seed consistently lead to significantly  higher losses across all metrics, highlighting the dominant influence of the seed on image generation outcomes.
    The asterisks **** correspond to p-values smaller than 0.0001.
    }
    \label{fig:eval:seed-significance}
\end{figure}

Statistical significance of observed differences between the SSDM and DSSM conditions was assessed using the Wilcoxon signed-rank test (sometimes referred to as Wilcoxon T-test)~\cite{DBLP:journals/technometrics/Groggel00/Wilcoxon-signed-rank-test}, a non-parametric statistical hypothesis test suitable for paired samples. This test evaluates whether differences between paired observations significantly deviate from zero without assuming normal distribution. The distributions of evaluated metrics under both conditions (SSDM and DSSM) are visualized in \Cref{fig:eval:seed-significance}, while \Cref{tab:example-images-ssdm-dssm} illustrates the same effect on three random example images.

\noindent\bheading{Results}
For Latent-MSE, images generated with different seeds (DSSM) demonstrated substantially higher losses (median approximately 1) compared to those generated with the same seed but varying modifiers (SSDM, median approximately 0.25). This difference was statistically highly significant, with a p-value of less than 0.0001. %
Similarly, the LPIPS metric exhibited  significantly greater losses for the DSSM condition (median approximately 0.7) compared to SSDM (median approximately 0.4), with high  statistical significance. %
The semantic loss measured via CLIP embeddings, while overlapping in the standard divergence, also showed significantly higher values in the DSSM condition (median approximately 0.19) compared to SSDM (median approximately 0.12), again with high statistical significance. %

\noindent\bheading{Discussion}
The findings consistently demonstrate that differences in the seed exert a stronger influence on the perceptual and semantic characteristics of generated images than variations in the stylistic modifiers alone. While CLIP loss performs best of all three, prior research by Williams \emph{et al.}~\cite{DBLP:journals/corr/abs-2408-06502/DiscreteOptimizersForPromptRecover} highlights its limitations in capturing perceptual variability as experienced by human observers.
This outcome highlights the critical role that the seed plays in defining the fundamental characteristics of images generated by diffusion models.

\subsection{Approximating the Seed: Is Near Enough Good Enough?}

\begin{figure}[t]
    \centering
    \includegraphics[width=1\linewidth]{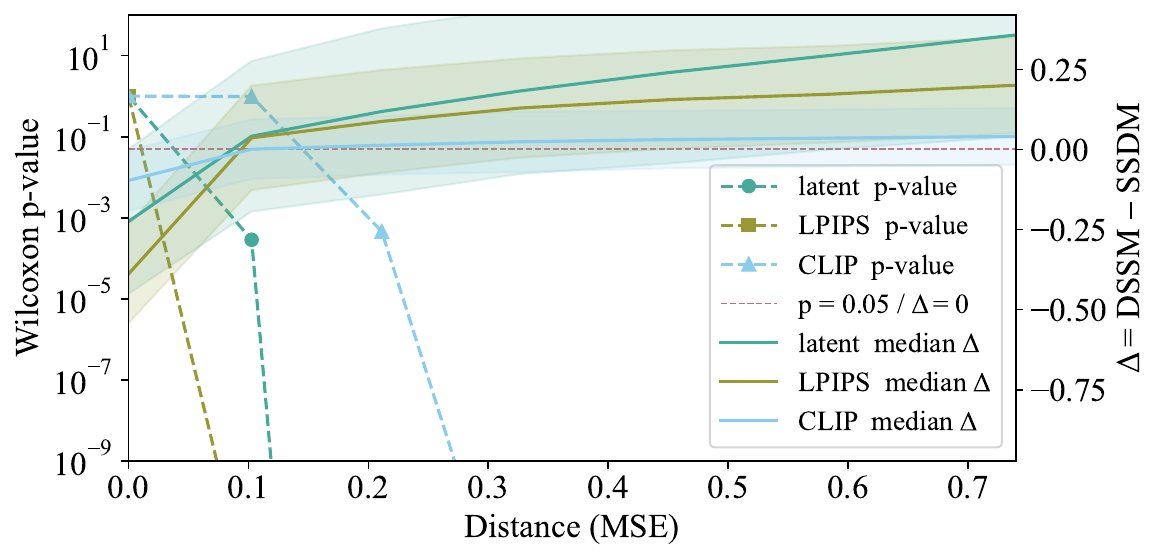}
    \caption{Impact of increasing noise approximation error (MSE) on the statistical significance of seed vs. modifier influence. Solid lines indicate the median difference \(\Delta = DSSM - SSDM\), while dashed lines indicate Wilcoxon signed-rank test p-values for each metric.
    }
    \label{fig:eval:noise-proximity-effect}
\end{figure}

The first experiment shows that the original seed and, consequently, the initial noise are crucial for accurate online prompt recovery. This naturally raises the question of whether a sufficiently close approximation might suffice to shift the focus from the seed to the stylistic modifiers. Thus, it is important to quantify how closely an approximation must match the true initial noise for modifiers to dominate image variation. Once this threshold is known, it is valuable to explore whether practical strategies exist to achieve such close approximations. Therefore, we evaluate several approaches to approximate the initial noise vector.

\noindent\bheading{Experiment Design}
This experiment consists of two parts. First, we replicate the setup from \Cref{sec:eval:seed-vs-modifier} but vary the distance from the original noise vector systematically. Initially, at a distance of 0, we essentially reassess the experiment from \Cref{sec:eval:seed-vs-modifier}. In subsequent rounds, we gradually increase the distance between the original noise vector and the approximation (quantified via mean squared error, MSE), while repeating the measurements. We record how the increased distance affects the relative influence of seed versus modifier changes on the generated images. These results are shown in \Cref{fig:eval:noise-proximity-effect}.

In the second part, we evaluate several concrete approaches to approximating the initial noise vector. We selected 1,000 random prompts and measured how closely each method’s reconstructed noise vectors matched the true initial noise vector used for image generation. This is quantified using mean squared error (MSE) in latent space. The methods we compared are:
\begin{itemize}
    \item \textbf{Gradient-based optimization}: Initialized with random noise, we optimized the approximated initial noise latent $\epsilon_s$ using the Adam optimizer for 500 iterations.
    The loss function combines the MSE between the current candidate $\epsilon_s$ and target latent $z_0$ vectors, variance regularization for distributional similarity, and a random-noise regularization term.
    \item \textbf{Original image latent vector}: Directly using the latent vector derived from the target image as a naive baseline.
    \item \textbf{Second-best seed}: The latent vector closest to the target latent found by brute-forcing 100,000 alternative seeds.
    \item \textbf{Random seed}: An entirely unrelated random seed.
\end{itemize}
The results from this second experiment are summarized in \Cref{tab:eval:different-techniques-to-approximate-original-noise}. Additionally, an illustrative example of how incremental changes to the input noise vector $\epsilon_s$ influences a generated image is provided in the Appendix (\Cref{tab:appendix:incremental-noise-change}).

\begin{table}[t]
    \centering
    \begin{threeparttable}
    \caption{How close can different techniques get to the initial noise $z_T$ given the final image $z_0$.}
    \label{tab:eval:different-techniques-to-approximate-original-noise}
    \begin{tabularx}{0.9\linewidth}{Xl}
    \toprule
    Method & Distance \\
    \midrule
         Optimization & $1.00 \pm 0.02$ \\
         Original Image & $1.98 \pm 0.21$ \\
         2nd-best seed (100000 other seeds) & $1.98 \pm 0.01$ \\
         Random Seed & $2.00 \pm 0.01$ \\ \bottomrule
    \end{tabularx}
    \end{threeparttable}
\end{table}

\noindent\bheading{Results}
The results for the first part of the experiment are presented in \Cref{fig:eval:noise-proximity-effect}.  
For all loss metrics, i.e., Latent-MSE, LPIPS, and CLIP, the Wilcoxon signed-rank test p-values (dashed lines) fall below the conventional significance threshold of \(p = 0.05\) already at very small distances. This suggests that even minimal deviations from the original noise introduce statistically significant differences. Of the three metrics, only CLIP maintains statistical robustness up to an approximation error of about 0.15.
Solid lines represent the median difference \(\Delta = DSSM - SSDM \), which represent the changes upon altering either the seed or a modifier, compared to the original image.
When \(\Delta < 0\), modifiers dominate the loss,  when \(\Delta > 0\), seeds dominate. As the noise approximation error grows, all metrics cross the \(\Delta = 0\) threshold around an MSE of approximately 0.1, indicating that beyond this point, seed differences become more influential than modifier differences. This finding aligns with the insights from \Cref{sec:eval:seed-vs-modifier}.

For the second part of the experiment, shown in \Cref{tab:eval:different-techniques-to-approximate-original-noise}, optimization-based approximation achieves the closest reconstruction with an average distance of \(1.00 \pm 0.02\). This result significantly outperforms naive baselines such as using the original image latent (\(1.98 \pm 0.21\)), the second-best seed (\(1.98 \pm 0.01\)), and a completely random seed initialization (\(2.00 \pm 0.01\)).

\noindent\bheading{Discussion}
Our findings show that one must be very close to the original noise, with a maximum MSE distance of 0.1, for modifiers to be more important than the seed.
Although optimization-based methods substantially outperform naive baselines, they remain far from the exact initial noise. The observed distances suggest that optimization alone does not produce close enough approximations to reliably shift the focus from seed variations to modifier differences. Therefore, knowledge of the precise seed is critical for numerical online prompt stealing, which emphasizes the importance of accurate seed recovery strategies for practical attacks.

\subsection{Seed stealing}
\label{sec:eval:seed-stealing}

The first two experiments demonstrate that the original noise distribution can only be generated if the original seed is known. Thus, this experiment evaluates the extent to which seed values can be recovered through brute-force attacks. We conducted an empirical experiment designed to measure both feasibility and reliability. Specifically, we examine how easily attackers could reconstruct the original random seed used in the Stable Diffusion 3.5 reference implementation, given its constrained seed space.

\noindent\bheading{Experiment Design}
To evaluate the feasibility of seed recovery, we conducted a comprehensive experiment leveraging a dataset of prompts from previous work by Shen \emph{et al.}~\cite{DBLP:conf/uss/ShenQ0024/UsenixImagePromptStealing}. Specifically, we randomly selected 1,000 prompts from the dataset and generated corresponding images using randomly chosen seed values within the allowed range (0 to 100,000), as in the reference implementation of Stable Diffusion 3.5~\cite{stabilityai_sd3_infer}.

We applied a brute-force seed recovery strategy to each generated image, systematically enumerating all 100,000 possible seeds in a simple Python implementation. In this implementation, we simply calculate the loss between the two latent variables. For each candidate seed $s$, we computed the Mean Squared Error (MSE) loss between the latent representation obtained from the candidate seed's noise vector $\epsilon_s$ and the latent representation of the original target image $z_0$. The seed yielding the lowest loss was identified as the predicted seed. Additionally, we compared the loss values of the best candidate seed against the second-best seed and the average loss across all candidate seeds to validate the robustness of our approach.

\begin{figure}[t]
    \centering
    \includegraphics[width=0.3\linewidth]{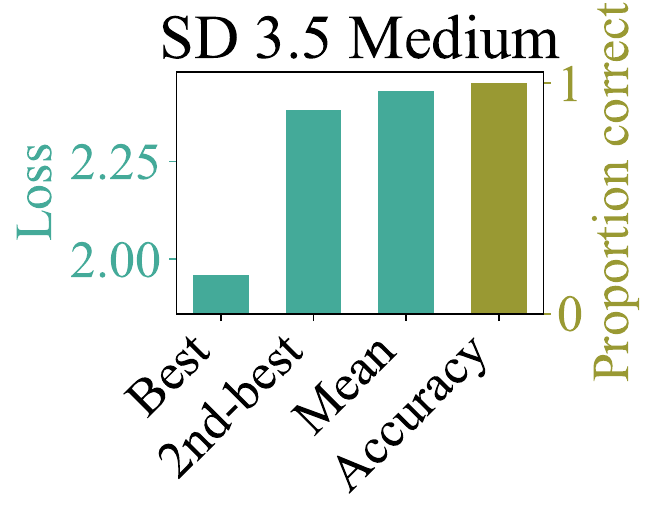}
    \includegraphics[width=0.3\linewidth]{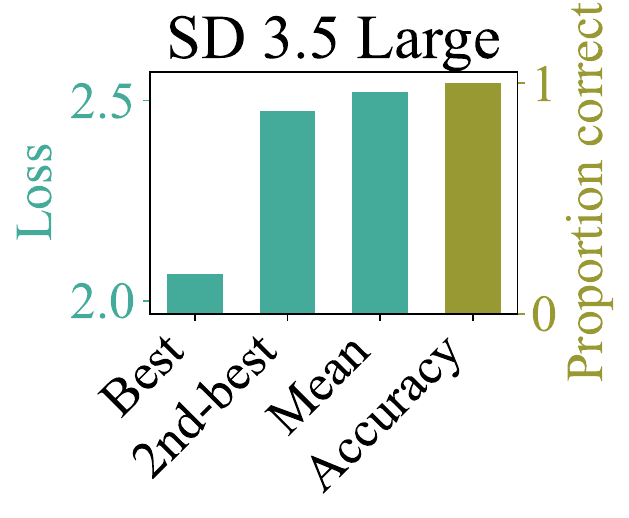}
    \includegraphics[width=0.3\linewidth]{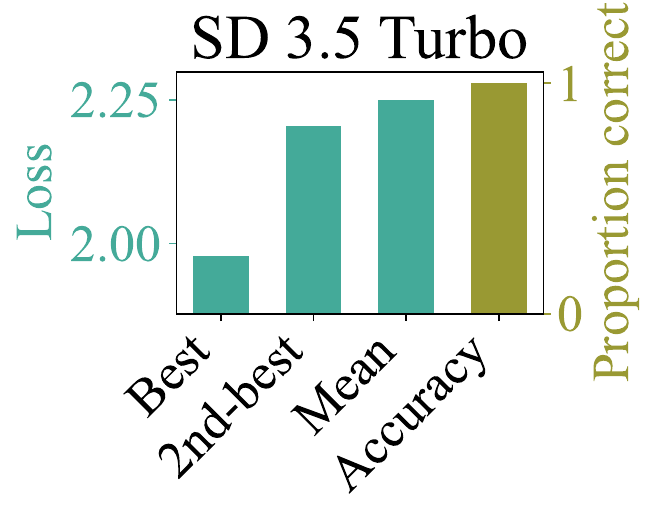}
    \caption{The MSE-loss between the initial noise $\epsilon_s$ and the target image $z_0$. The correct seed consistently shows significantly lower loss values compared to the second-best seed and the average loss. Accuracy in seed identification is 100\% across all models.}
    \label{fig:seed_recovery_performance}
\end{figure}

\noindent\bheading{Results}
Our experiment demonstrated perfect accuracy (100\%) in recovering the correct seed across all 1,000 trials. Figure~\ref{fig:seed_recovery_performance} highlights the clear distinction between the best seed (correct seed) and other candidate seeds. The best seed consistently produced a significantly lower loss compared to the second-best candidate and the overall average. Specifically, the average loss of the best seed across trials was substantially smaller compared to the second-best seed and the overall mean loss value. On average, our prototype required approximately 85.2 seconds to recover the seed for each image.

\noindent\bheading{Discussion}
The results unequivocally indicate that the initial noise seed is strongly embedded in the generated image, enabling reliable seed recovery through brute-force enumeration. Due to the stark difference in loss values between the correct seed and incorrect seeds, attackers can easily validate seed correctness by comparing candidate seeds against a sample of randomly chosen incorrect seeds to establish a baseline loss value. This significantly undermines the security of generative image models relying on limited seed spaces and suggests the critical need for larger, more secure seed ranges.

\subsection{Comparison with State-of-the-Art}
\label{sec:eval:comparison-with-state-of-the-art}
Following the initial experiments, we determined that the initial seed is necessary for loss functions to indicate how close an image is to another, while accounting for the influence of the seed. Additionally, we discovered that the seed can be identified and recovered. In the following, we evaluate the effectiveness of prompt stealing under the assumption that the seed is known.

To evaluate the performance of our proposed approach, we compare \ourapproach{} (PP) with three well-known existing methods for prompt extraction. One offline method and two online methods. The offline method is Prompt Stealer by Shen \emph{et al.}~\cite{DBLP:conf/uss/ShenQ0024/UsenixImagePromptStealing}. It was chosen, as it represents state-of-the-art in offline prompt stealing. For online methods, we compare CLIP Interrogator~\cite{clip_interrogator} and P2HP~\cite{DBLP:conf/cvpr/MahajanRYS24/PromptingHardOrHardly}. We chose CLIP Interrogator because it is a widely-used method for reverse-engineering prompts from images, also used as a baseline in related work~\cite{DBLP:conf/uss/ShenQ0024/UsenixImagePromptStealing}. Additionally, P2HP was selected as it represents the current state-of-the-art numerical optimization-based prompt-stealing method, surpassing methods such as PEZ~\cite{DBLP:conf/nips/WenJKGGG23/PEZ}.

We evaluate two scenarios: \textit{Known Subject}, where the main subject is provided, and \textit{Unknown Subject}, where the subject must also be inferred. The rationale behind this separation is that inferring the subject from an image is generally easier, whereas accurately capturing stylistic modifiers represents the core challenge. Drawing an analogy to art, the subject of the \textit{Mona Lisa} could simply be described as \textit{a portrait of a woman}, but the modifiers define the unique style and artistic value of the painting.

\noindent\bheading{Experiment Design}
We evaluated each technique using their default configurations on images from the test split of the Shen \emph{et al.} dataset~\cite{DBLP:conf/uss/ShenQ0024/UsenixImagePromptStealing}. For fairness, each target image was generated using the specific Stable Diffusion model for which the respective approach was developed. To avoid any unfair advantage from leaked parameters, we did not reuse the original seed during the evaluation. Instead, we fix a single evaluation seed that is shared across all methods and renders. Each method outputs a candidate prompt, which we then rendered using the shared fixed seed and compared to the original prompt rendered with that seed. Qualitative panels in \Cref{tab:eval:images-from-prompt-recover} illustrate this protocol. 
This evaluation step is repeated five times with different seeds, and the average value is reported to ensure robustness.
Due to the longer runtime of certain methods, we restricted the experiment to 100 images. In order to adapt P2HP to allow for arbitrary subjects as a hint, we modified its second optimization step by prepending the correct subject directly. For \ourapproach{}, modifiers were sourced from the train split of the Shen \emph{et al.} dataset, applying a minimum usage threshold of 1\% to filter out infrequently used modifiers and reduce noise, while the subject was generated using their provided captioning model.

\begin{figure}[t]
    \centering
    \includegraphics[width=1\linewidth]{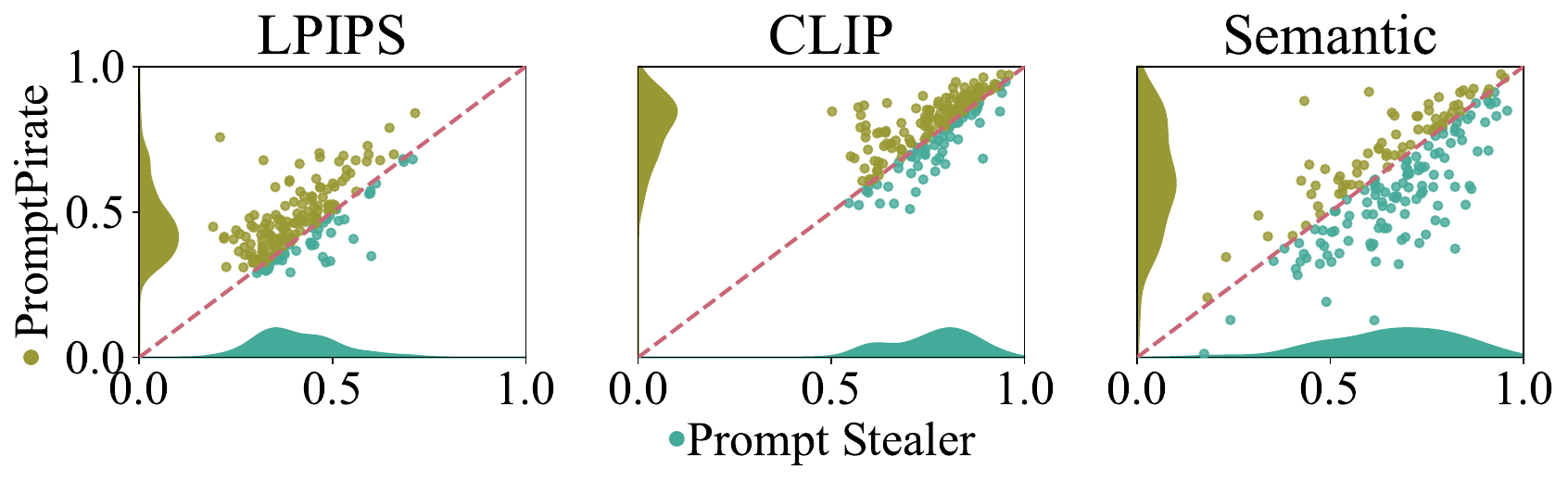}
    \caption{A detailed comparison between the best two approaches. All data points were plotted with the y-value as the performance of \ourapproach{} and the x-value as the performance of Prompt Stealer. Therefore, whenever a dot is above the diagonal line, our approach is better, and vice versa. The distribution of dots is shown on the respective axis.}
    \label{fig:eval:us-vs-usenix}
\end{figure}

To evaluate the techniques comprehensively, we used the following metrics recommended by prior work~\cite{DBLP:conf/uss/ShenQ0024/UsenixImagePromptStealing, DBLP:conf/cvpr/MahajanRYS24/PromptingHardOrHardly}:
\begin{itemize}
    \item \textbf{Perceptual Similarity:} Given the target prompt and the recovered prompt, we compute the LPIPS similarity between the original and reconstructed images, reflecting perceptual similarity from a human viewpoint.

    \item \textbf{Content Similarity:} The content similarity is the cosine similarity between the CLIP embeddings of the original and reconstructed images, capturing visual content coherence.

    \item \textbf{Semantic Similarity:} This metric measures the cosine similarity between CLIP embeddings of the original text prompt and the recovered text prompt, assessing how closely the inferred textual content matches the target semantics.

\end{itemize} 
The results of these metrics are summarized in \Cref{tab:eval:performance-comparison}. A detailed comparison between the two best-performing techniques is provided in \Cref{fig:eval:us-vs-usenix}. Additionally, we present qualitative comparisons for four randomly selected images from the dataset in \Cref{tab:eval:images-from-prompt-recover} (although the samples were chosen at random, we needed to test two seeds to avoid images containing nudity or religious blasphemy). The original and recovered prompts for each random image is shown in the Appendix in \Cref{tab:appendix:sample-images-prompts}.

\begin{table}[t]
    \centering
    \caption{Performance comparison of multiple prompt stealing techniques. Best mean values are bold.}
    \label{tab:eval:performance-comparison}

    \begin{threeparttable}
        \begin{tabularx}{1\linewidth}{X|ccc}
\toprule 
     \multicolumn{4}{c}{Known Subject} \\ \midrule
    & LPIPS $\uparrow$ & CLIP $\uparrow$ & Semantic $\uparrow$  \\

Prompt Stealer & 0.47 $\pm$ 0.14& 0.80  $\pm$ 0.13& \textbf{0.77  $\pm$ 0.14} 
 \\
CLIP Interrogator 
& 0.40 $\pm$ 0.10& 0.78  $\pm$ 0.11& 0.58  $\pm$ 0.14 
 \\
  P2HP  & 0.43 $\pm$ 0.11& 0.77  $\pm$ 0.11& 0.48  $\pm$ 0.22 \\
     \ourapproach{}  & \textbf{0.52 $\pm$ 0.14}& \textbf{0.83  $\pm$ 0.11}& 0.71  $\pm$ 0.18 

\\
\midrule
    \multicolumn{4}{c}{Unknown Subject}  \\ \midrule
    & LPIPS $\uparrow$ & CLIP $\uparrow$ & Semantic $\uparrow$  \\
Prompt Stealer & 0.37  $\pm$ 0.09& 0.71  $\pm$ 0.11& \textbf{0.61 $\pm$ 0.15} 
\\
CLIP Interrogator & 0.37  $\pm$ 0.07& \textbf{0.75  $\pm$ 0.10}& 0.42 $\pm$ 0.11 \\
    P2HP & 0.35  $\pm$ 0.07& 0.66  $\pm$ 0.11& 0.16 $\pm$ 0.17 
 \\
     \ourapproach{} & \textbf{0.40  $\pm$ 0.08}& \textbf{0.75  $\pm$ 0.11}& 0.51 $\pm$ 0.14  

\\

\bottomrule
\end{tabularx}
    \end{threeparttable}
\end{table}

\begin{table}[t]
    \centering

    \begin{threeparttable}
        \caption{Visual comparison between the evaluated techniques, Prompt Stealer(PS), CLIP Interrogator (CI), P2HP and \ourapproach{} (PP), on random sample images. }
        \label{tab:eval:images-from-prompt-recover}
        \begin{tabular}{l|cccccc} \toprule

\multicolumn{5}{c}{Known Subject}  \\ \midrule

Original  & PS & CI & P2HP & \textbf{PP} \\ \midrule

            \includegraphics[width=0.15\linewidth]{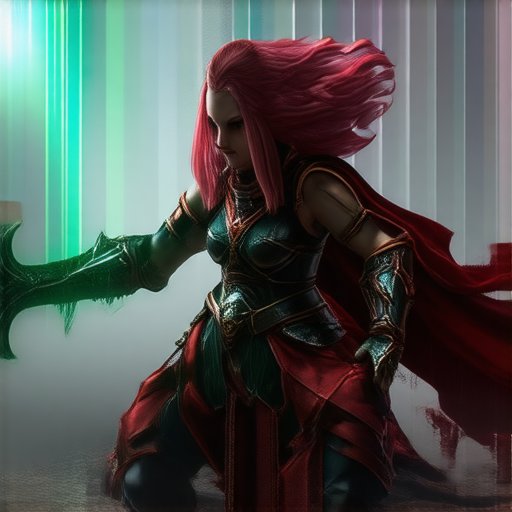} & 
            \includegraphics[width=0.15\linewidth]{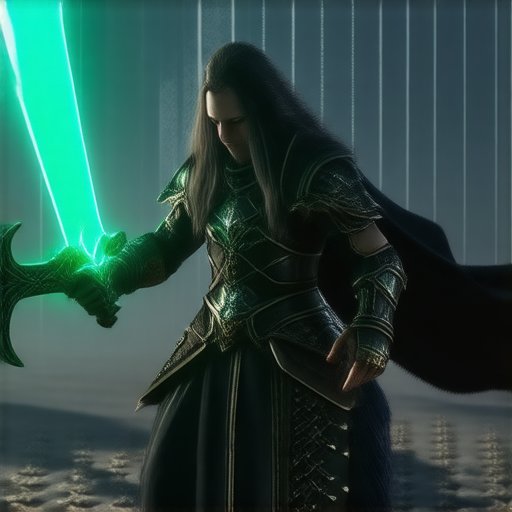} &
            \includegraphics[width=0.15\linewidth]{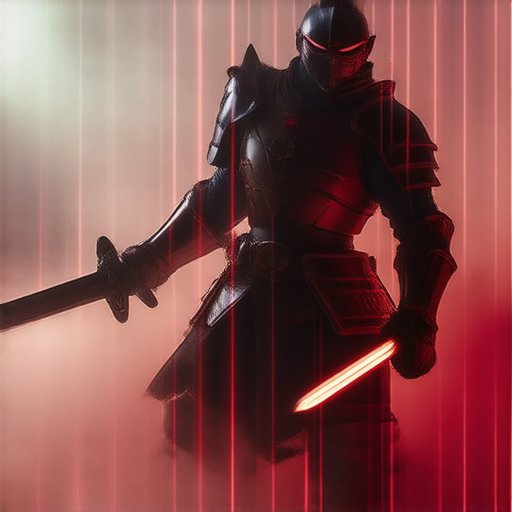}  &
            \includegraphics[width=0.15\linewidth]{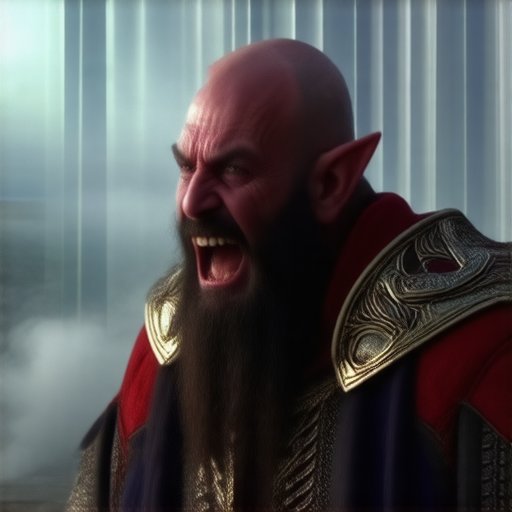}  &
            \includegraphics[width=0.15\linewidth]{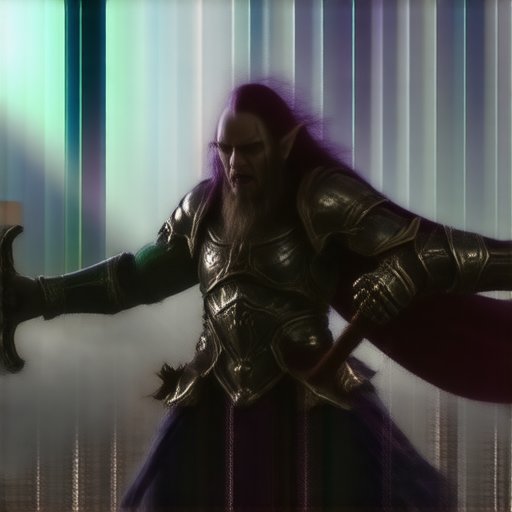}   \\

            \includegraphics[width=0.15\linewidth]{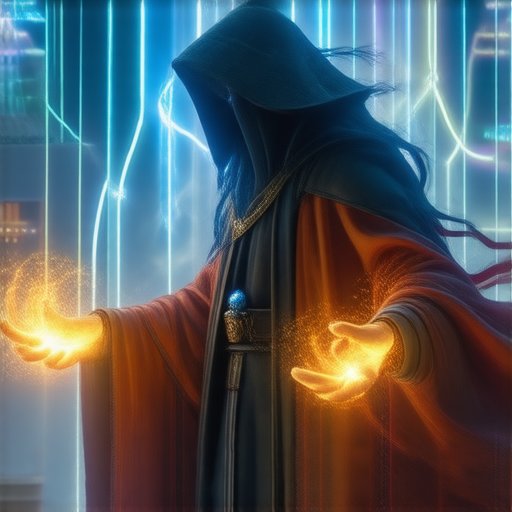} & 
            \includegraphics[width=0.15\linewidth]{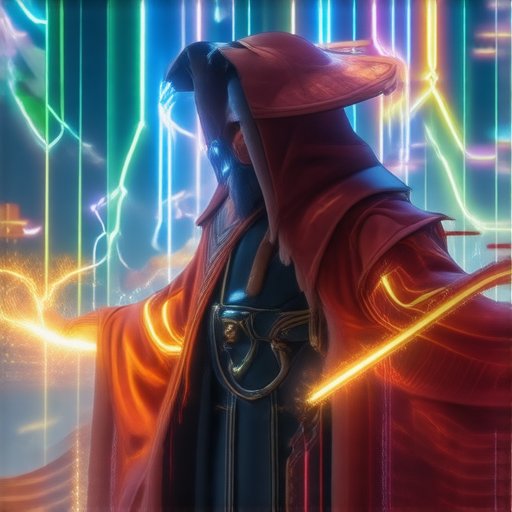} &
            \includegraphics[width=0.15\linewidth]{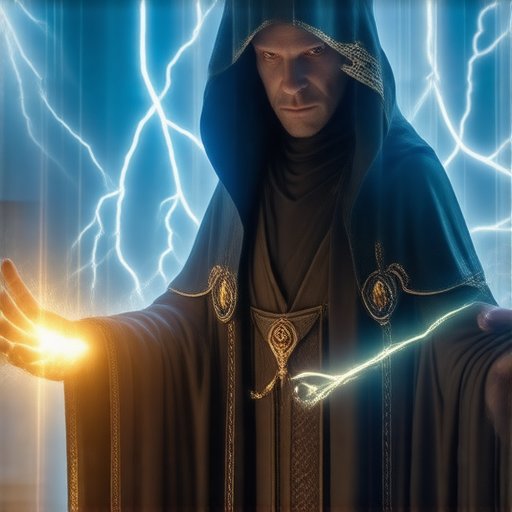}  &
            \includegraphics[width=0.15\linewidth]{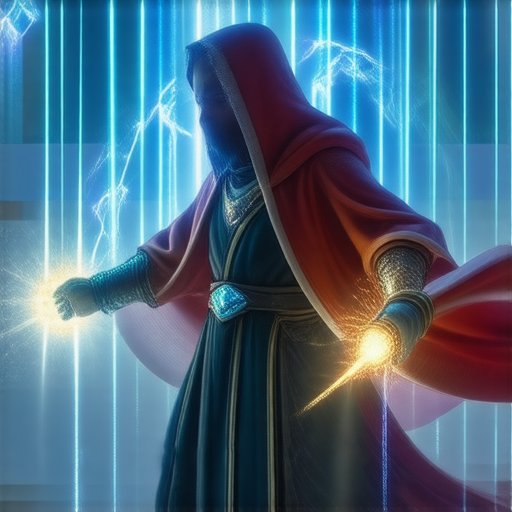}  &
            \includegraphics[width=0.15\linewidth]{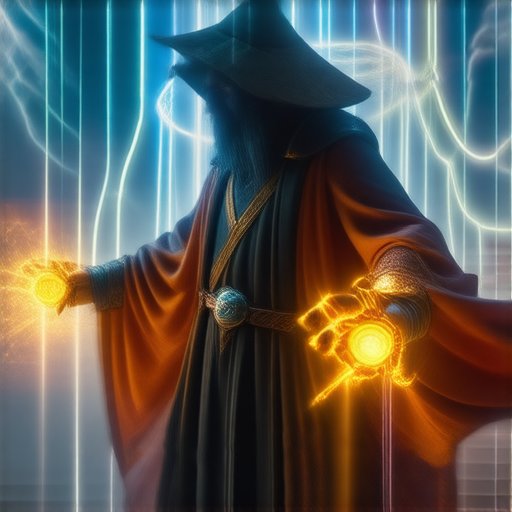}   \\

            \includegraphics[width=0.15\linewidth]{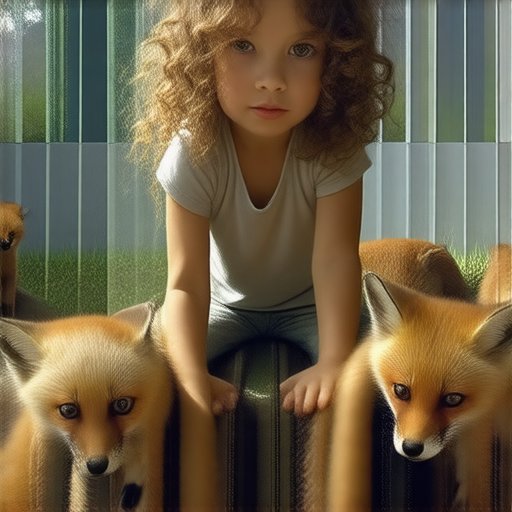} & 
            \includegraphics[width=0.15\linewidth]{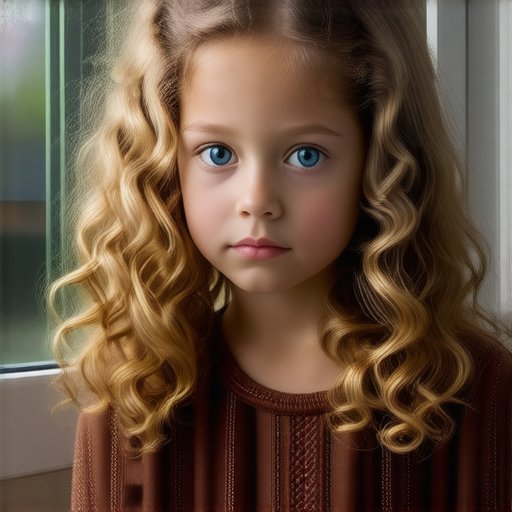} &
            \includegraphics[width=0.15\linewidth]{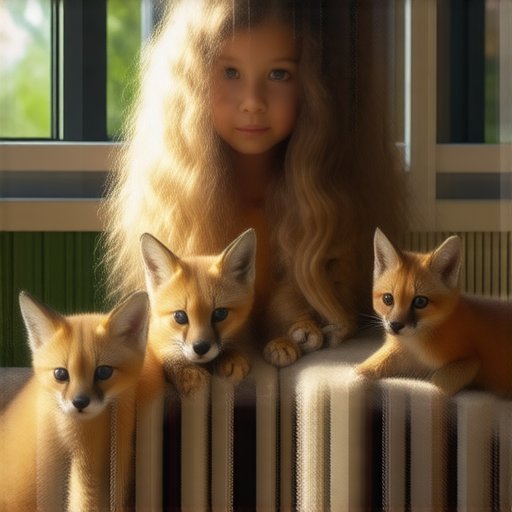}  &
            \includegraphics[width=0.15\linewidth]{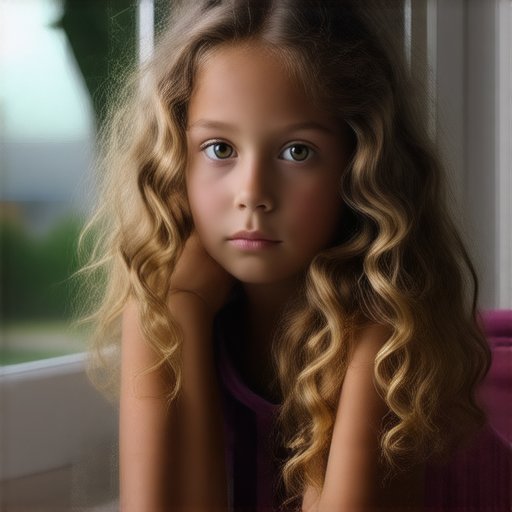}  &
            \includegraphics[width=0.15\linewidth]{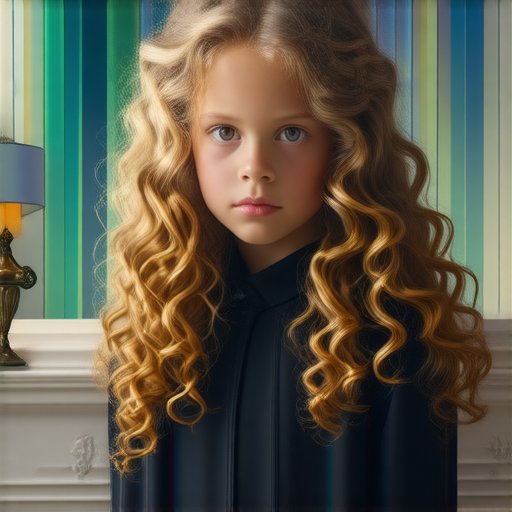}   \\

            \includegraphics[width=0.15\linewidth]{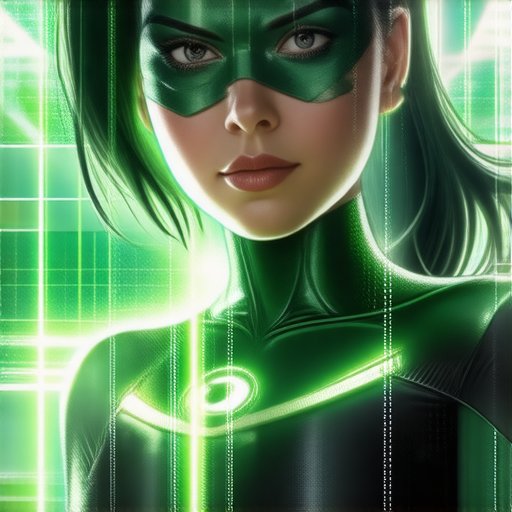} & 
            \includegraphics[width=0.15\linewidth]{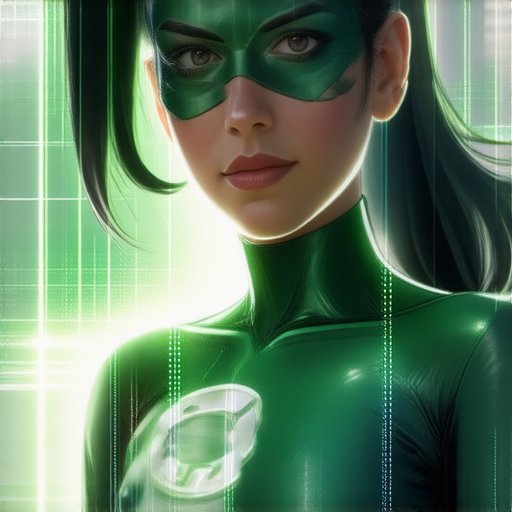} &
            \includegraphics[width=0.15\linewidth]{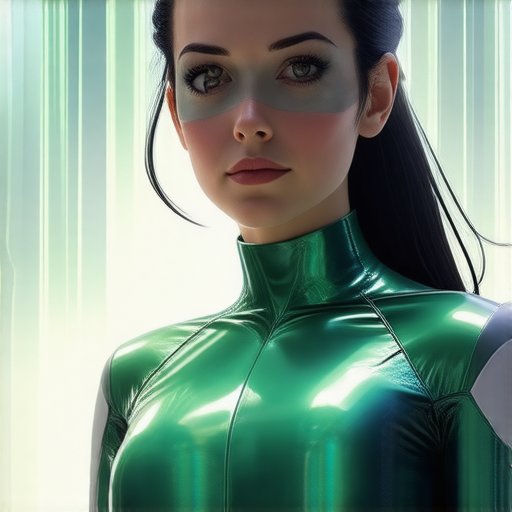}  &
            \includegraphics[width=0.15\linewidth]{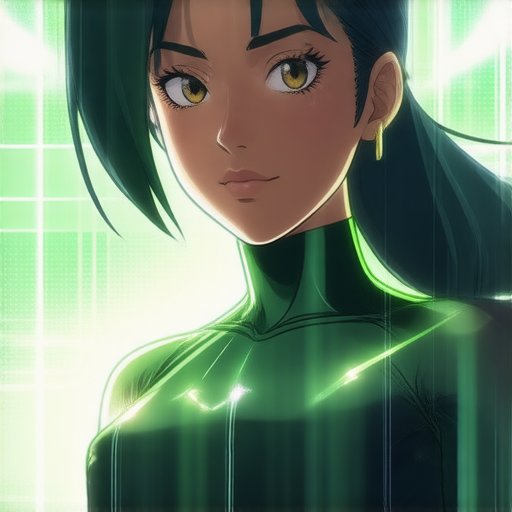}  &
            \includegraphics[width=0.15\linewidth]{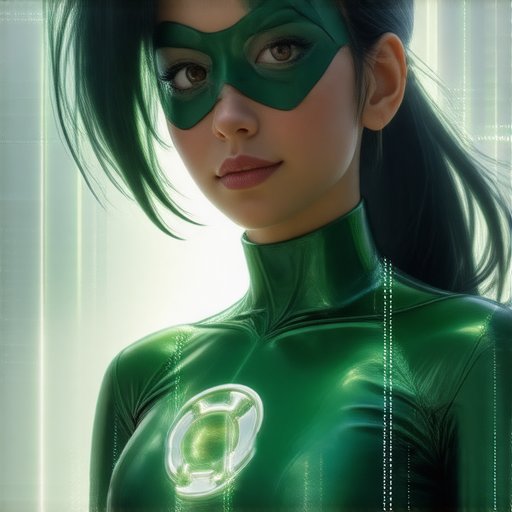}  \\

\bottomrule 
\toprule
        \multicolumn{5}{c}{Unknown Subject}  \\ \midrule
        
        Original & PS & CI & P2HP & \textbf{PP} \\ \midrule

            \includegraphics[width=0.15\linewidth]{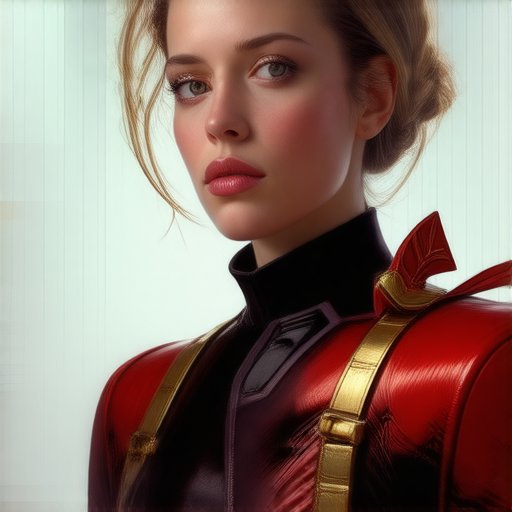} & 
            \includegraphics[width=0.15\linewidth]{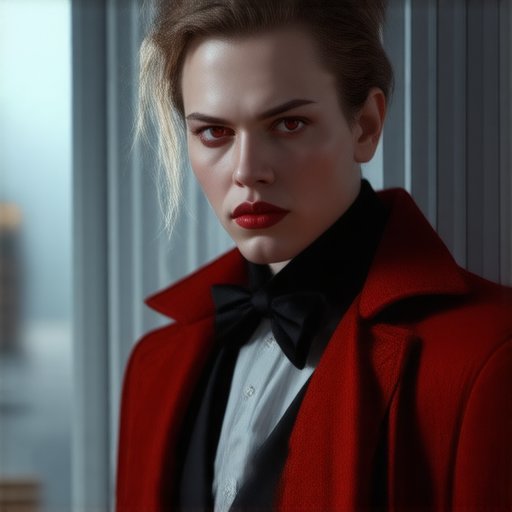} & 
            \includegraphics[width=0.15\linewidth]{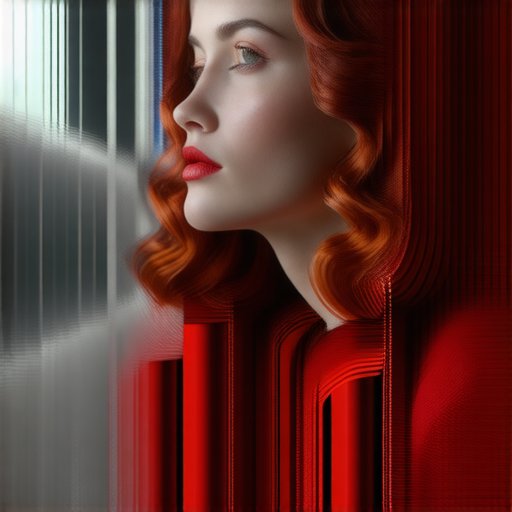}  &
            \includegraphics[width=0.15\linewidth]{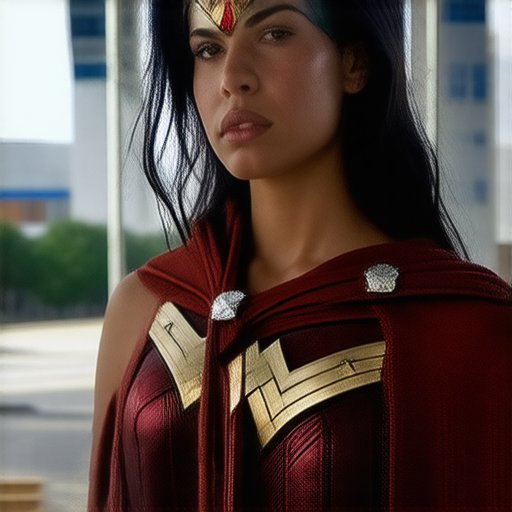}  &
            \includegraphics[width=0.15\linewidth]{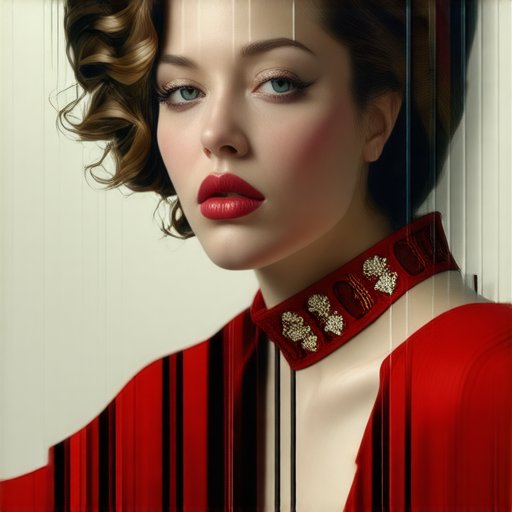}   \\

            \includegraphics[width=0.15\linewidth]{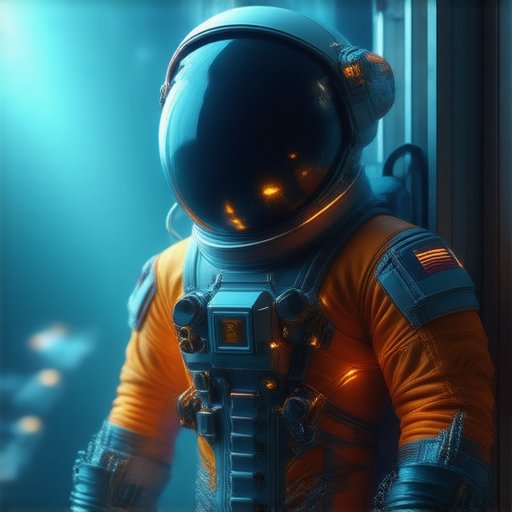} & 
            \includegraphics[width=0.15\linewidth]{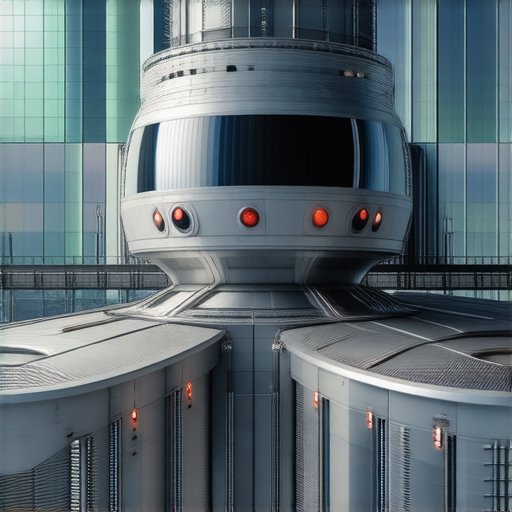} & 
            \includegraphics[width=0.15\linewidth]{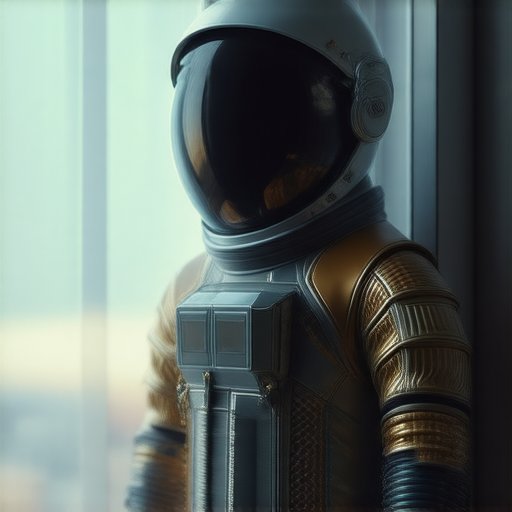}  &
            \includegraphics[width=0.15\linewidth]{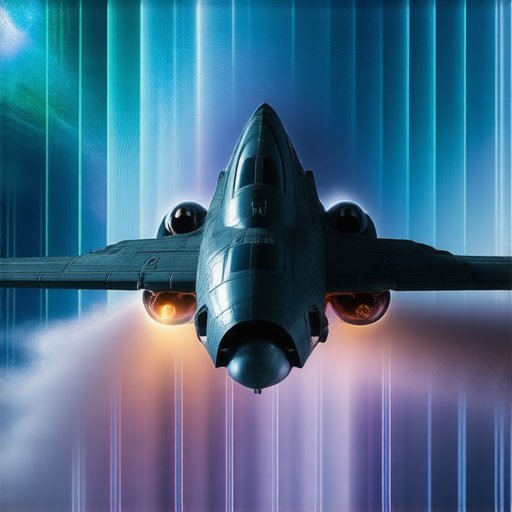}  &
            \includegraphics[width=0.15\linewidth]{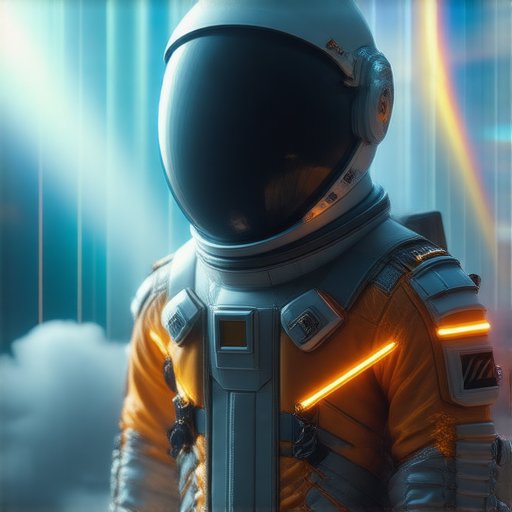}   \\

            \includegraphics[width=0.15\linewidth]{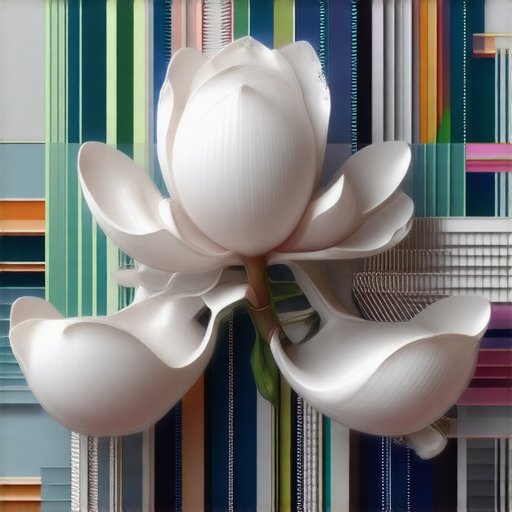} & 
            \includegraphics[width=0.15\linewidth]{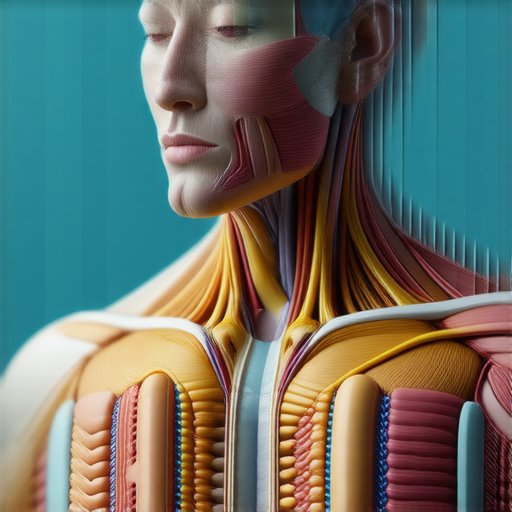} & 
            \includegraphics[width=0.15\linewidth]{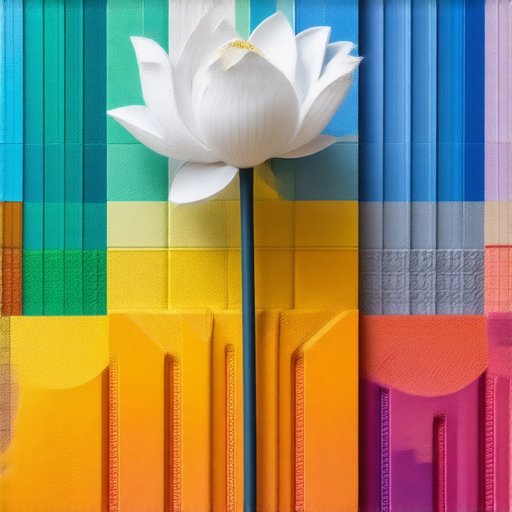}  &
            \includegraphics[width=0.15\linewidth]{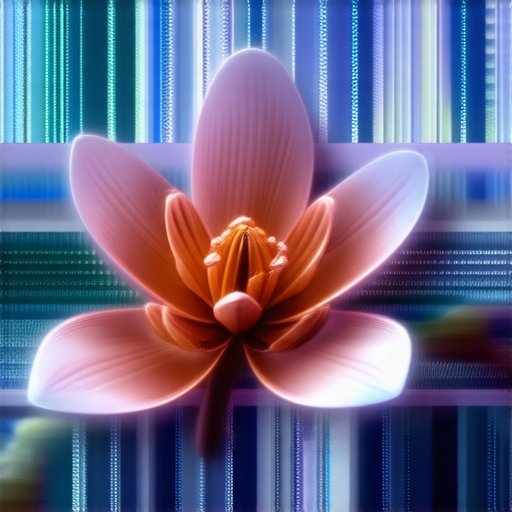}  &
            \includegraphics[width=0.15\linewidth]{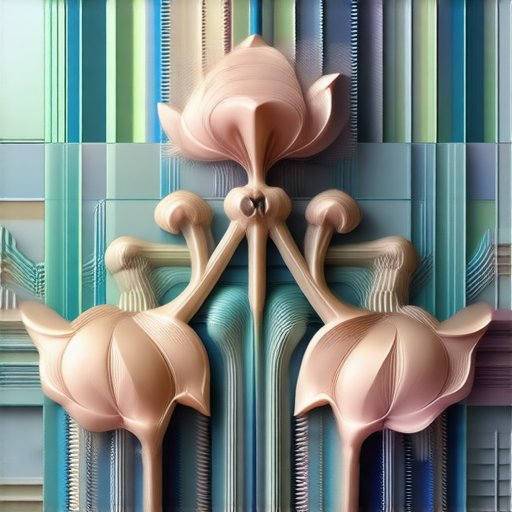}   \\

            \includegraphics[width=0.15\linewidth]{images/comparison_image_samples/original_4_666_False.jpg} & 
            \includegraphics[width=0.15\linewidth]{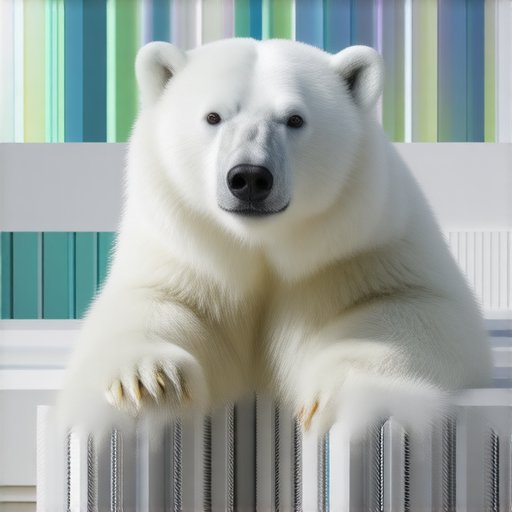} & 
            \includegraphics[width=0.15\linewidth]{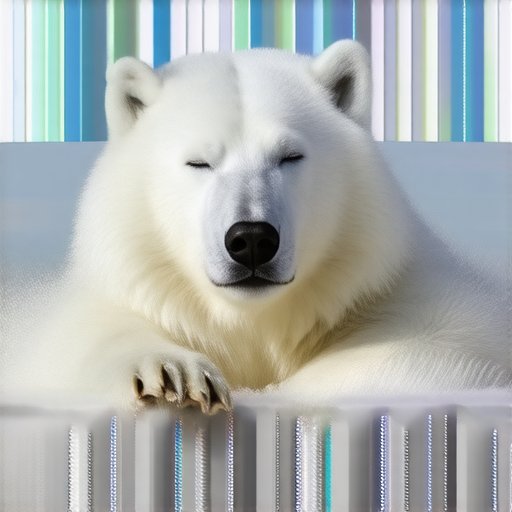}  &
            \includegraphics[width=0.15\linewidth]{images/comparison_image_samples/P2HP_4_1_False.jpg}  &
            \includegraphics[width=0.15\linewidth]{images/comparison_image_samples/us_4_1_False.jpg}   \\
            
            \bottomrule
            
        \end{tabular}
    \end{threeparttable}
    
\end{table}

\noindent\bheading{Results} 
As shown in \Cref{tab:eval:performance-comparison}, \ourapproach{} consistently matches or surpasses existing state-of-the-art methods across visual similarity metrics, and also surpasses all online methods in text recovery metrics. Prompt Stealer (Offline) consistently achieves the highest semantic similarity, which aligns with its goal of generating prompts closely resembling the original text. This confirms its strength in capturing semantic meaning. However, it underperforms in visual similarity metrics (e.g., LPIPS and CLIP), suggesting that its outputs, while semantically rich, do not reliably guide the generation of visually faithful reconstructions, which is also the attacker's goal. 
In the \textit{Known Subject} scenario, \ourapproach{} achieves the best perceptual similarity (LPIPS: $0.52 \pm 0.14$), representing an $11\%$ relative increase compared to the next-best baseline (Prompt Stealer at $0.47$). This is accompanied by a CLIP score increase from $0.80$ (Prompt Stealer) and $0.78$ (CLIP Interrogator) to $0.83$. However, Prompt Stealer attains the highest semantic similarity ($0.77$), about $9\%$ higher than \ourapproach{} ($0.71$). This is likely because Prompt Stealer tends to produce overly verbose or literal prompts, which, while semantically rich, do not always translate into visually accurate or faithful reconstructions, highlighting a tradeoff between semantic and visual alignment.

In the more challenging \textit{Unknown Subject} setting, \ourapproach{} again achieves the highest perceptual similarity (LPIPS: $0.40$), and matches CLIP Interrogator in CLIP similarity ($0.75$). Prompt Stealer maintains the top semantic score ($0.61$), but its generated visuals lack fidelity, suggesting that high semantic similarity in this case may be due to overfitted language patterns, as the evaluation dataset is from the same source as the dataset used to train their model. Thus, it is of a similar distribution to this evaluation dataset.

Qualitative examples in \Cref{tab:eval:images-from-prompt-recover} reflect these trends. In the \textit{Known Subject} rows, all methods produce images more faithful to the original compared to the unknown-subject case. However, P2HP exhibits clear failures: it misses critical content like the complete figure in row one or the sparkle effects in row two. Although it performs closer to \ourapproach{} or Prompt Stealer in row three, it still omits key attributes like the mask and style in row four. CLIP Interrogator occasionally identifies objects (e.g., the dogs in row three) but often misses visual structure, such as poses or colors. Prompt Stealer adds hallucinated elements, like a laser sword in row one and struggles with structure, e.g., misrepresenting capes or omitting the dogs in row two. In contrast, \ourapproach{} consistently maintains high visual fidelity.

Similar observations hold in the \textit{Unknown Subject} panel. P2HP fails to recover key subjects like the astronaut (row 2) or the sleeping dog (row 4). Prompt Stealer also misses the astronaut and floral details. CLIP Interrogator identifies many objects but lacks accuracy in visual positioning and expression. \ourapproach{} offers stronger spatial and stylistic consistency, better capturing nuanced cues such as the dog's pose and the woman’s facial expression.

We further analyze PromptPirate's and Prompt Stealer's relative performance in \Cref{fig:eval:us-vs-usenix}, where each data point represents the performance for a specific prompt. Here, Prompt Stealer's performance is represented on the x-axis, while \ourapproach{}'s performance is represented on the y-axis. Points above the diagonal indicate prompts for which \ourapproach{} outperforms Prompt Stealer, whereas points below the diagonal indicate the opposite.

From this analysis, it is evident that \ourapproach{} outperforms Prompt Stealer across both visual metrics (LPIPS and CLIP), with most data points positioned above the diagonal. Particularly for LPIPS, which correlates strongly with human visual perception, \ourapproach{} exhibits clear superiority. In contrast, Prompt Stealer tends to perform better on the semantic metric, although \ourapproach{} also shows strong performance for many prompts, demonstrating competitive results even in semantic comparisons.

In terms of runtime, measured on an NVIDIA A100 GPU, notable differences emerge among the evaluated methods. Prompt Stealer, as offline approach, requires $21.20$ seconds to process the entire dataset of $100$ images. However, this measurement excludes the offline training phase and the generation of training data. In comparison, the remaining methods function entirely online, without any offline training component. Among these online methods, \ourapproach{} requires approximately $62.36$ minutes per image, CLIP Interrogator takes roughly $1.02$ minutes per image, and P2HP has the highest computational demand, needing $140.44$ minutes per image.

\bheading{Discussion}
Our evaluation highlights a key finding: incorporating seed recovery substantially improves the effectiveness of online prompt stealing. While the offline method (Prompt Stealer) achieves higher semantic similarity scores, \ourapproach{} consistently produces prompts that result in images more visually aligned with the original.
This distinction is crucial. As Williams \emph{et al.}~\cite{DBLP:journals/cviu/CroitoruHIS24} point out, multiple prompts can yield indistinguishable images, meaning that recovering the exact phrasing of the original prompt is not always necessary or even uniquely determined. Instead, the core objective in prompt recovery should be to identify a prompt that faithfully reconstructs the visual content. From this perspective, \ourapproach{} provides a more practically valuable solution: it discovers prompts that are not only semantically plausible but also effective at reproducing the original image. This makes it better suited for real-world prompt recovery tasks where both textual and visual fidelity are essential.

\bheading{Ablation Study} In an ablation study with $N{=}50$ prompts, we rerun \ourapproach{} with an incorrect (non--stolen) seed while holding all other components fixed. This single change yields an average $5\%$ drop across all reported metrics, establishing that the correct seed is essential for a successful online prompt–stealing attack. Consistent with \Cref{sec:eval:seed-vs-modifier}, the seed fixes the stochastic generation path so that fitness differences are attributable to the modifier tokens rather than sampling noise. %

\section{Case Study: CivitAI}
\label{sec:case-study-civitai}

To evaluate the real-world impact and practical relevance of our seed-recovery approach, we conducted an empirical case study. We used publicly available images from CivitAI, a widely used online platform that hosts user-generated images from various Stable Diffusion models. Specifically, we collected a dataset of Stable Diffusion 3.5 images along with their openly shared generation metadata, including prompts, seeds, and sampler settings.

This case study enables us to investigate the theoretical effectiveness of our proposed attack and its implications for privacy and intellectual property security. By analyzing the distribution of seeds chosen by real users and verifying our method's ability to precisely recover these seeds, we gain valuable insights into the genuine risk exposure creators face when sharing content online.

\subsection{Seed Identification}
Our goal is to evaluate the practicality and accuracy of identifying generation seeds from images produced by the Large, Medium, and Turbo variants of the Stable Diffusion 3.5 models. This assessment provides critical insights into the real-world applicability and privacy implications of our seed recovery method beyond lab conditions.

\noindent\bheading{Experiment Design} We conducted our experiments using publicly accessible images on CivitAI, focusing specifically on the Stable Diffusion versions 3.5 Large, Medium, and Turbo. First, we retrieved all images explicitly associated with the Medium and Turbo variants. Due to the considerably larger number of available images for the Large variant, we randomly selected 1,000 samples to ensure a representative yet manageable subset. After  removing images lacking explicit model version details or precise seed metadata, the final curated dataset consisted of 895 images, i.e., 640 images from Stable Diffusion 3.5 Large, 179 images from Medium, and 76 images from Turbo.
We performed seed identification by attempting to recover the correct seed for each image from a candidate pool consisting of the original seed and 100,000 randomly selected distractor seeds.

\noindent\bheading{Results}
Our seed identification method demonstrated high accuracy across all model variants, with 95\% of correct seed identification. For the Large model, we successfully identified the correct seed in 94\% of cases. Performance improved with the Medium model, achieving 98\% accuracy. Remarkably, the method achieved perfect accuracy (100\%) on the Turbo model.

\noindent\bheading{Discussion} Our method effectively identifies seeds in realistic, community-generated datasets, highlighting substantial real-world privacy risks. 
Only $5\%$ of seeds could not be identified. One possible explanation for those $5\%$ is post-generation editing of images, which is common practice. Minor imperfections might be manually corrected or refined using InFill networks that redraw specific areas. Another explanation could be the employment of an alternative, possibly proprietary, random number generator or another distinct method for generating the initial noise $\epsilon_s$. A third hypothesis, that seed detection might be impossible in these cases, appears less likely given our empirical findings in \Cref{sec:eval:seed-stealing}, where we reliably identified all correct seeds of real world prompts.
Thus, we consider the inability to detect seeds in this third cluster unlikely, leaning instead toward image editing or alternative RNG methods as more plausible explanations.

\begin{figure}[t]
    \centering
    \includegraphics[width=1\linewidth]{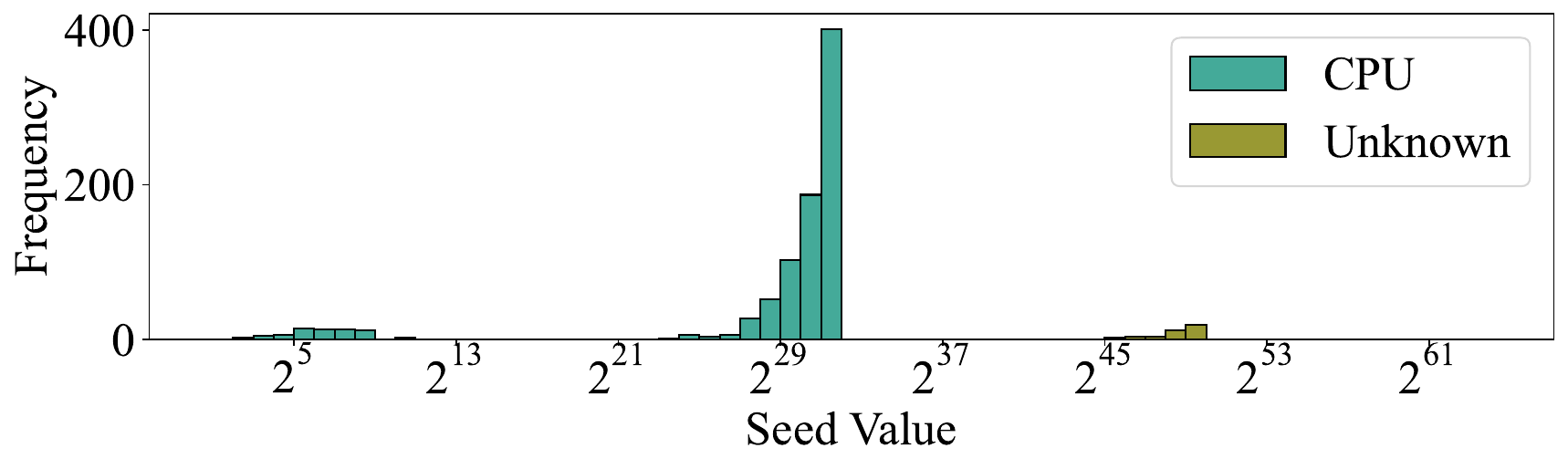}
    \caption{Effective seed size on CivitAI}

    \label{fig:eval:CivitAI-seed-histogram}
\end{figure}

\subsection{Seed Distribution}
\label{sec:eval:civitai-seed-distribution}
To better understand the practical implications of our seed-recovery attack, we examined the distribution of seed values used for image generation in the real world images from CivitAI. This analysis helps identify common practices and potential vulnerabilities associated with seed selection patterns among content creators.

\noindent\bheading{Experiment Design} 
We analyzed the complete dataset of 895 images collected from CivitAI, with a specific focus on the distribution of their associated seed values. The seeds were extracted directly from the metadata provided with each image.
As explained in \Cref{sec:rng-vulnerabilities}, the initial noise generation depends only on the least significant 32 bit of the seed, when generated on the CPU. Thus, we only consider these parts of the seed in SeedSnitch.
We plotted all the seeds and categorized these values to visually and quantitatively assess their distribution patterns.

\noindent\bheading{Results}
As shown in \Cref{fig:eval:CivitAI-seed-histogram}, the seed values from the dataset exhibit a clear clustering pattern across three distinct ranges. The largest and most prominent cluster spans from $2^{29}$ to $2^{32}$, representing users who generated seeds using the MT19937 RNG algorithm on CPUs. These CPU-based users also contributed to a smaller, distinct cluster of much lower seed values ranging from approximately $2^5$ to $2^8$. In total, 95\% of seeds are effectively in a 32-bit range. A third cluster, presumably generated using an unknown RNG algorithm spans the range from $2^{42}$ to $2^{50}$.

\noindent\bheading{Discussion} 
The identified clustering patterns provide valuable insights into users' practical seed-generation behaviors. The MT19937-based cluster's ($2^{29}$–$2^{32}$) dominant presence underscores a widespread reliance on predictable CPU-generated seeds, highlighting a significant real-world vulnerability. The smaller, low-value seed cluster (0–512) indicates manual seed selection and further exacerbates privacy risks due to the extremely limited search space required for potential brute-force attacks. In summary $95\%$ of seeds could be brute forced.

\subsection{Seed Recovery via 32bit Brute-Force}
\label{sec:eval:CivitAI:seed-brute-force}

To empirically validate the practical feasibility and real-world impact of brute-forcing seeds using the MT19937 algorithm, we conduct a case study on  publicly available images from CivitAI.

\noindent\bheading{Experiment Design}
We selected 50 random images from CivitAI, where we knew from the previous experiment that the initial noise was generated using the CPU. We then downloaded the corresponding JPG files, thus mimicking the behavior of a potential attacker.
Each of these images publicly includes metadata specifying the associated seed. However, for our study, we deliberately disregarded the provided seeds, treating them as unknown. Our goal is to assess the feasibility and efficiency of brute-forcing the seed solely based on the generated image, while having the ground truth to calculate the performance.

We implemented a Python-based prototype that exhaustively searches the complete 32-bit seed space supported by PyTorch’s MT19937 algorithm. To reduce computational costs compared to the full-vector comparison described in \Cref{sec:eval:seed-stealing}, we introduced a two-stage filtering approach. The entire 32-bit seed range is divided into chunks of size $2^{16}$, enabling parallel processing. Within each chunk, seeds are initially evaluated using only the first $2^{13}$ entries of the noise vector. This reduces the computational load by approximately 87.5\% for a standard 512×512 pixel image, while still preserving accuracy. We found that further reduction (e.g., using fewer entries) notably degraded accuracy.
The top $k=2^{13}$ seed candidates from the first stage are re-evaluated with higher precision using the first $2^{15}$ values of the noise vector. Empirically, $2^{15}$ values are sufficient for perfect identification. This two-stage approach balances computational efficiency and accuracy. Adding a third intermediate filtering stage did not improve accuracy but slowed computation. Decreasing $k$ below $2^{13}$ decreased accuracy. 
The brute-force process was executed on two Intel Xeon Gold 6342 CPUs (2.80 GHz), using multi-threading capabilities to maximize efficiency.

\noindent\bheading{Results}
Our experiment successfully recovered the correct seeds for all fifty images from CivitAI with 100\% accuracy. The brute-force enumeration of the entire 32-bit MT19937 seed space required approximately 140 minutes per image on the utilized CPU setup.

\noindent\bheading{Discussion}
The 100\% success rate and practical runtime observed underscore a severe and realistic vulnerability within the Stable Diffusion ecosystem, particularly concerning the predictability and limited randomness provided by the MT19937 algorithm when used with a 32-bit seed space. This case study confirms that seed brute-forcing attacks are not merely theoretical but can be efficiently executed in realistic, community-generated scenarios. Additionally, we believe that switching from our basic, multi-threaded Python implementation to a more sophisticated one would result in a significant increase in speed.
Given the feasibility demonstrated, this emphasizes the critical necessity for transitioning toward more secure random number generation approaches.

\section{Mitigation and Security Discussion}

The vulnerabilities exploited by SeedSnitch and identified in the Stable Diffusion model arise from the limited 32-bit seed space commonly used in current implementations. This constrained seed space significantly facilitates brute-force attacks. However, the weaknesses associated with PyTorch's internal random number generator are not new. Therefore, PyTorch's developers previously introduced the cryptographically secure pseudorandom number generators (CSPRNG) extension~\cite{csprng2020}. This extension can operate in two distinct modes:

    \noindent\textbf{Cryptographically secure randomness:} Randomness is derived directly from secure sources such as \textit{/dev/urandom} and subsequently encrypted using AES. This mode, however, does not support seed-setting, which is critical in image generation processes. The ability to set a seed enables consistent initialization of noise, crucial for iterative prompt engineering tasks. Although it is theoretically possible to save and reuse noise directly, this practice is not used in current image generation frameworks or user-interfaces.

    \noindent\textbf{Seed-supported secure randomness:} This mode utilizes the MT19937 algorithm, whose output is subsequently encrypted via AES. This approach still relies on a 32-bit seed space, inherently restricting security. Therefore, merely adopting the CSPRNG extension does not sufficiently address the vulnerability.

To effectively mitigate these issues while preserving reproducibility via seed-setting, we propose adopting a cryptographically secure random number generator (e.g., ChaCha20~\cite{DBLP:journals/rfc/rfc7539/ChaCha}) with a significantly expanded seed range, such as 256 bits, ensuring quantum resistance. This approach would necessitate only minor code modifications and introduces minimal computational overhead while robustly preventing brute-force and cryptographic attacks. Replacing PyTorch’s default RNG with ChaCha20 increases the noise-sampling step by roughly $8\times$. However, because noise sampling constitutes only $0.005\%$ (SD 3.5 Large) to $0.037\%$ (SD 3.5 Turbo) of end-to-end generation time, the resulting overall overhead is negligible in practice. Unlike in related-work settings where RNG throughput can dominate system cost, e.g., in differential privacy pipelines~\cite{DBLP:conf/uss/DahiyaS024}, randomness generation here is a vanishingly small fraction of total runtime.

\section{Related Work}
\noindent\bheading{Offline} Early approaches to prompt extraction in text-to-image diffusion models often treat prompts as fixed conditioning vectors, overlooking the role of the stochastic seed in image generation. PromptStealer, proposed by Shen \emph{et al.}~\cite{DBLP:conf/uss/ShenQ0024/UsenixImagePromptStealing}, uses supervised learning by combining a captioning model to recover the subject and a multi-label classifier to infer style attributes. However, these components require retraining for each version of the diffusion model. Similarly, Reverse Stable Diffusion by Croitoru \textit{et al.}~\cite{DBLP:journals/cviu/CroitoruHIS24} learns to recover prompts by first predicting their sentence embeddings from generated images, and then using a second model to decode these embeddings back into text. These learning-based methods demand large paired datasets and retraining for each new model version. In contrast, \ourapproach{} is online and therefore model-agnostic. By exploiting a CWE-339 vulnerability that limits the seed space, we first brute-force the seed and then perform seed-aware optimization to recover the prompt efficiently.

\noindent\bheading{Online} Orthogonal to offline methods, \textit{online} prompt recovery approaches (which optimize prompts directly during image generation) have also emerged. Mahajan \textit{et al.}'s P2HP~\cite{DBLP:conf/cvpr/MahajanRYS24/PromptingHardOrHardly} optimizes token embeddings during the late stages of denoising and subsequently maps them back to valid vocabulary tokens, effectively recovering \emph{hard prompts}, i.e., discrete text inputs rather than continuous embeddings. Williams \textit{et al.}~\cite{DBLP:journals/corr/abs-2408-06502/DiscreteOptimizersForPromptRecover} later evaluated a range of discrete optimizers (e.g., PEZ~\cite{DBLP:conf/nips/WenJKGGG23/PEZ}) and found that CLIP-based objectives often overfit to text embeddings, occasionally allowing even simple captioning baselines to outperform more sophisticated optimizers. Complementing these insights, \ourapproach{} shows that, once the correct seed is recovered, even a lightweight genetic optimizer can reliably outperform existing online methods.

\noindent\bheading{Seed Recovery} A parallel line of research investigates the semantic impact of seeds. Xu \emph{et al.}~\cite{DBLP:conf/wacv/XuZS25/Golden_Seed} identified that certain “golden” seeds consistently yield higher-fidelity images and demonstrated that seed classification can achieve near-perfect accuracy within a limited candidate set of 1,024 seeds. While their work positions seeds as quality-control parameters, \ourapproach{} uniquely frames the seed as a \emph{security primitive}. Importantly, we show that the seed recovery classifier proposed by Xu \textit{et al.} is an unnecessary abstraction. In Section~\ref{sec:seed-recovery-formal}, we demonstrate that the exact seed can be recovered \emph{without} relying on any learned model. Moreover, while their method is constrained to predicting among 1,024 candidates, our approach operates over the \emph{entire} seed space used by the diffusion model. To the best of our knowledge, this is the first work to systematically investigate and establish the impact of seed information on prompt stealing.

\noindent\bheading{Diffusion Security} Beyond prompt extraction, existing security research has identified other privacy vulnerabilities in generative models. Carlini \textit{et al.}~\cite{DBLP:conf/uss/CarliniHNJSTBIW23} extracted copyrighted imagery directly from large diffusion checkpoints. Duan \textit{et al.}~\cite{DBLP:conf/icml/DuanK0SX23} demonstrated diffusion models' susceptibility to membership-inference attacks, and earlier work by Hilprecht \textit{et al.}~\cite{DBLP:journals/popets/HilprechtHB19} reported similar vulnerabilities in GANs and VAEs. These studies primarily highlight leakage of training data, whereas \ourapproach{} demonstrates complementary leakage, specifically the exposure of the generation seed and prompt.

In summary, prior studies either (i) ignore seed recovery and depend heavily on supervised training, (ii) invert prompts under unrealistic assumptions without seed knowledge, or (iii) investigate seed semantics without linking them to prompt confidentiality. 

\ourapproach{} closes the gap by combining practical seed extraction with an online, seed-aware prompt stealing method. This uncovers a previously overlooked CWE-339 vulnerability that enables real-world prompt-stealing attacks.

\section{Conclusion}
In this paper, we have demonstrated that effective optimization-based prompt stealing from diffusion models fundamentally depends on accurately recovering the initial random seed used during image generation. Through rigorous empirical analysis, we established that standard similarity metrics (Latent-MSE, LPIPS, and CLIP) exhibit significantly higher sensitivity to variations in initial seeds than to prompt modifiers. This emphasizes the important role of the seed in determining image characteristics and its impact on the feasibility of prompt extraction attacks.

We identified and exploited a CWE-339 vulnerability in multiple popular implementations of Stable Diffusion-based image generators. Our analysis revealed that constrained seed ranges and inadequately robust pseudo-random number generation significantly reduce seed complexity, facilitating practical brute-force attacks. Thus, we proposed SeedSnitch, a robust and efficient %
approach specifically crafted for practical seed extraction. %
Through real-world validation with publicly available images, %
we demonstrated that SeedSnitch is highly effective, achieving successful brute-force recovery of 95\% of seeds within 140 minutes. This underscores alarming 
security practices prevalent in current image generation workflows.

Building upon successfully recovered seeds, we introduced \ourapproach{}, a genetic algorithm explicitly designed to extract prompt modifiers from generated images. %
\ourapproach{} consistently outperforms previous state-of-the-art methods, producing significantly more accurate and visually similar reconstructions of images generated from secret prompts.  

Finally, we emphasized the %
necessity of adopting secure random number generation practices. We recommend integrating cryptographically robust RNG algorithms into diffusion model implementations to mitigate the identified vulnerabilities effectively. Our findings and proposed mitigations significantly advance the understanding of security threats inherent in diffusion models and offer  practical strategies for safeguarding intellectual property embedded within generated images.

\section*{Acknowledgements}
This work was supported by the Federal Ministry of Research, Technology and Space (BMFTR) through the projects \href{https://samsmart.de/en/}{\textit{SAM Smart}} and \href{https://anomed.de/}{\textit{AnoMed}}.

\bibliographystyle{plain}
\bibliography{References}

\section{Appendix}

\begin{table*}[h]
    \centering

{\smaller[1.5] \begin{tabularx}{\linewidth}{clX}
\toprule

\multicolumn{3}{c}{Known Subject}  \\ \midrule

\#img & Type & Prompt \\ \midrule
1 & Original & evil elves fight evil dwarves, composition, trending on artstation, cinematic \\

 & Prompt Stealer & evil elves fight evil dwarves, artstation, 8k, intricate, 4k, epic lighting, award winning photograph \\

 & CLIP I. & evil elves fight evil dwarves, red - toned mist, evil knight, shutterstock contest winner, runescape, eldar samurai,  solo performance unreal engine, gearing up for battle, final fantasy xiv, cinematic rule of thirds, marketing photo, hyper realistic ”, 500px, mazinger, politics, dramatic ”, gurren lagan, blades, roleplay \\

 & P2HP & evil elves fight evil dwarves, cinematic diablo diablo vg diablo steal giants yelling \\

 & \ourapproach{} & evil elves fight evil dwarves, epic composition, cinematic, octane \\
\midrule

2 & Original & a powerful mage surrounded by electrical force field, mixed media, digital art, trending on artstation, 8k, epic composition, highly detailed, AAA graphics \\

 & Prompt Stealer & a powerful mage surrounded by electrical force field, artstation, highly detailed, concept art, 8k, sharp focus, digital painting, intricate, illustration, smooth, 4k, fantasy, cinematic, cinematic lighting, unreal engine, detailed, by ilya kuvshinov, cgsociety, beautiful, hyperrealistic, high quality, symmetrical, atmosphere, god rays, lens flare, glow, extreme details, magic, epic scale, majestic, advanced technology, holy \\

 & CLIP I. & a powerful mage surrounded by electrical force field, ww 1 sith sorcerer, like magic the gathering, beautiful avatar pictures, looks realistic, wearing mage robes, polycount contest winner, there is lightning, emperor secret society, ultrafine detail ” \\

 & P2HP & a powerful mage surrounded by electrical force field, dame conceptart incoming god rog elephreveals \\

 & PromptPirate & a powerful mage surrounded by electrical force field, hd, dynamic lighting, painted, volumetric lighting, 4k, epic, high contrast \\
\midrule
3 & Original & a seven - year old with long curly dirty blonde hair, blue eyes, tan skin a tee shirt and shorts, playing with foxes, painting by daniel gerhartz, alphonse mucha, bouguereau, detailed art, artstation  \\

 & Prompt Stealer & a seven - year old with long curly dirty blonde hair, artstation, highly detailed, digital painting, by craig mullins, by gaston bussiere, leyendecker, blue eyes \\

 & CLIP I. & a seven - year old with long curly dirty blonde hair, playing with foxes, enhanced photo, [ [ hyperrealistic ] ], with afro, indoor picture, many cute fluffy caracals, pov photo, sunlights, trio, beauty shot, version 3, short legs, reflecting light, cute!, smooth edges, yellow carpeted, inspired by Anna Füssli \\

 & P2HP & a seven - year old with long curly dirty blonde hair, tender fairies innocence wondering dorothy clemens stowe hove \\

 & PromptPirate & a seven - year old with long curly dirty blonde hair, peter mohrbacher, colorful, tom bagshaw, masterpiece, fine details, global illumination, elegant, full body, art nouveau, global illumination, peter mohrbacher, centered \\
\midrule

4 & Original & anime key visual of a beautiful young female green lantern!! intricate, green and black suit, glowing, powers, dc comics, cinematic, stunning, highly detailed, digital painting, artstation, smooth, hard focus, illustration, art by artgerm and greg rutkowski and alphonse mucha  \\

 & Prompt Stealer & anime key visual of a beautiful young female green lantern!! intricate, artstation, highly detailed, concept art, sharp focus, digital painting, intricate, illustration, smooth, elegant, by artgerm, by artgerm and greg rutkowski and alphonse mucha, by wlop, unreal engine, by ilya kuvshinov, 3d, by craig mullins, by makoto shinkai, fine details, anime, art station, stunning, perfect face, atmospheric lighting, by mucha, hard focus, leesha hannigan, fantasy illustration, katsuhiro otomo, movie poster, light and shadow effects, epic light \\

 & CLIP I. & anime key visual of a beautiful young female green lantern!! intricate, digital painting - n 5, catsuit, family photo, ultrafine detail ”, wearing! robes!! of silver, charicature, summer glau, beautiful avatar pictures, desktop wallpaper, wide shot!!!!!!, phoenix, piggy, minimalism, concept art-h 640, lisa, jets \\

 & P2HP & anime key visual of a beautiful young female green lantern!! intricate, uq razer female elephsuperheroes girl aphra \\

 & PromptPirate & anime key visual of a beautiful young female green lantern!! intricate, ruan jia, 8k, artstation hq, global illumination, tom bagshaw \\

\midrule  \midrule
\multicolumn{3}{c}{Unknown Subject}  \\ \midrule

1 & Original & scarlet johansson, deep focus, d \& d, fantasy, intricate, elegant, highly detailed, digital painting, artstation, concept art, matte, sharp focus, illustration, hearthstone, art by artgerm and greg rutkowski and alphonse mucha  \\

& Prompt Stealer & scarlett johansson as a vampire, artstation, highly detailed, concept art, 8k, sharp focus, digital painting, intricate, illustration, smooth, elegant, by greg rutkowski, fantasy, cinematic lighting, by artgerm and greg rutkowski and alphonse mucha, unreal engine, by ilya kuvshinov, by alphonse mucha, unreal engine 5, by rossdraws, global illumination, by rhads, by loish, by tom bagshaw, detailed and intricate environment, by makoto shinkai and lois van baarle, radiant light, pixiv, by stephen bliss, by artgerm and greg rutkowski, by ferdinand knab, hearthstone, by anna dittmann, by nikolay makovsky, textured skin \\

& CLIP I. & arafed image of a woman with red hair and a red hood, abstract fractal automaton, lipstick, 8 k detail, red curtain, fibonacci composition, fashion portrait photo, intricate art deco pasta designs, masterpiece ; behance hd, semiabstract, lacquered glass, close - up photo \\

& P2HP & !!!shawl, wonderwoman, vampiassassins, wounded, if, \\

& \ourapproach{} & a portrait of a beautiful scarlett johansson, d \& d, james jean, extremely detailed, photorealistic, alphonse mucha, beautiful, peter mohrbacher, artgerm \\

\midrule

 2 & Original & concept art by craig mullins astronaut in futuristic dark and empty spaceship underwater. infrared complex and hyperdetailed technical suit. mandelbulb fractal. reflection and dispersion materials. rays and dispersion of light. volumetric light. 5 0 mm, f / 3 2. noise film photo. flash photography. unreal engine 4, octane render. interstellar movie art  \\

& Prompt Stealer & a photorealistic dramatic hyperrealistic render of a retrofuturistic space station in a, artstation, 8k, octane render, 4k, cinematic, cinematic lighting, unreal engine, photorealistic, hyper detailed, render \\

& CLIP I. & there is a man in a space suit standing in front of a window, octane render dynamic lighting, muted stage effects, inspired by Russell Chatham, up close picture, the image is refined with uhd, stock photo, and the uncertainty\', fresnel effect, keyframe illustration, blurred space, in flight suit, mid shot photo \\

& P2HP & vortex, bubble, ethereal, cyberpunk, nebula, scifi, ship, \\

& \ourapproach{} & a man in a spacesuit, radiant light, moebius, centered, matte, ross tran, cinematic lighting, fantasy art, d\&d, detailed and intricate environment \\
\midrule

3 & Original & beautiful classical decorative ornament, anatomy, energy, geometry, magnolia, bones, petals, stems, fibonacci rhythm, artstation, art germ, wlop  \\

& Prompt Stealer & anatomy of the human body, artstation, highly detailed, 8k, intricate, highd, post - processing, fibonacci flow, acroteria, encarpus, large medium and small elements \\

& CLIP I. & there is a large white flower on a pole in front of a colorful wall, symmetrical complex fine detail, 3 d sculpture of carving marble, pastel vivid triad colors, jugendstil background, inspired by Zaha Hadid, magnolia, hyperrealistic image of x, palace of the chalice, holy light, lotus, centered close-up, maxim sukharev, holy \\

& P2HP & demonic, flower, flake, sculpting, eurusd, ), froze, \\

& \ourapproach{} & beautiful art nouveau style wall sculpture, greg rutkowski, highly detailed, resolution, detailed, high quality, ross tran, stunning, rossdraws, symmetrical, rossdraws, symmetrical, ultra detailed \\

\midrule

4 & Original & giant sleeping ( ( ( ( ( ( white canine ) ) ) ) ) ) cute creature creature in a tundra landscape, dramatic lighting, moody : : by michal karcz, daniel merriam, victo ngai and guillermo del toro : : ornate, dynamic, particulate, intricate, elegant, highly detailed, centered, artstation, smooth, sharp focus, octane render, 3 d  \\

& Prompt Stealer & a photorealistic dramatic hyperrealistic render of a cute white anthropomorphic polar bear, concept art, 8k, intricate, octane render, 4k, highd, photorealistic, etail, hyper detailed, god rays \\

& CLIP I. & there is a polar bear that is laying down on the snow, samoyed dog, sweet dreams, highly photographic render, wolfy nail, high res photo, enhanced photo, eyes closed, very sharp photo, a stock photo \\

& P2HP & creepy, woolly, titanmonster, winter, alien, wildlife, \\

& \ourapproach{} & a beautiful painting of a white wolf sleeping on a bed, dramatic lighting, detailed face, intricate details, makoto shinkai and lois van baarle, detailed face, detailed face, beeple \\

\bottomrule \\
 
\end{tabularx}}
    \caption{The original and stolen prompts for the images shown in \Cref{tab:eval:images-from-prompt-recover}.}
    \label{tab:appendix:sample-images-prompts}
\end{table*}

\begin{table*}[]
    \centering
    \begin{tabular}{r|ccccccc}

\raisebox{0.045\linewidth-0.5em}[0em][0em]{Image} & \includegraphics[width=0.09\linewidth]{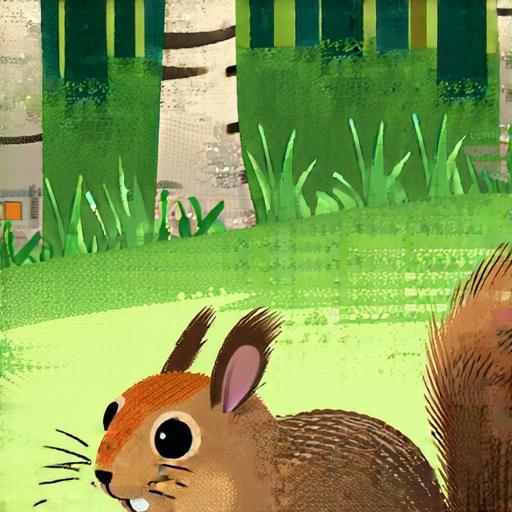} &\includegraphics[width=0.09\linewidth]{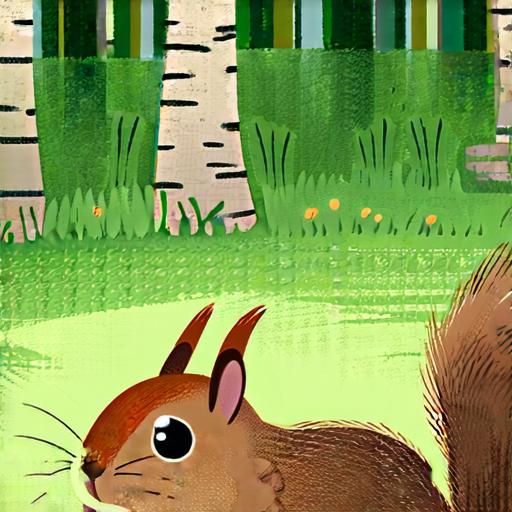} &\includegraphics[width=0.09\linewidth]{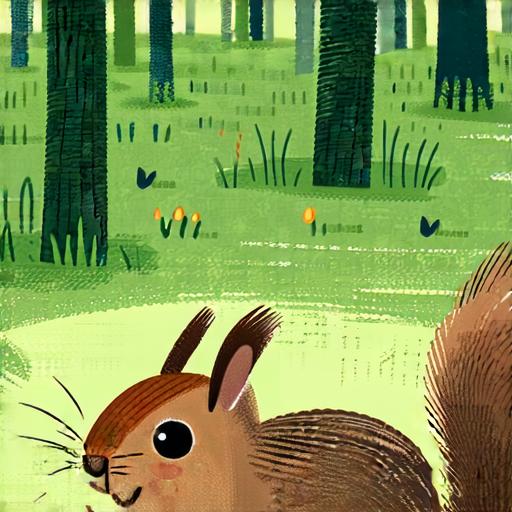} &\includegraphics[width=0.09\linewidth]{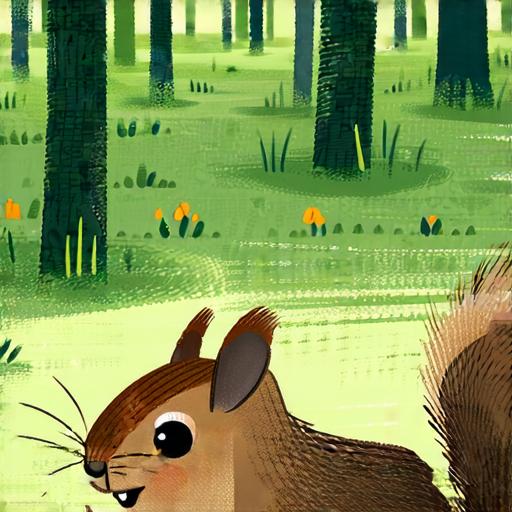} &\includegraphics[width=0.09\linewidth]{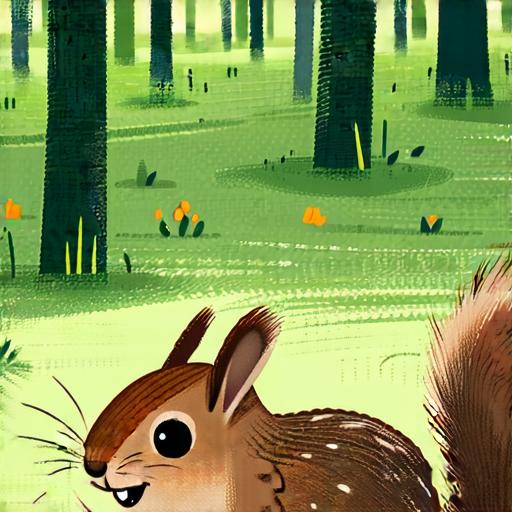} &\includegraphics[width=0.09\linewidth]{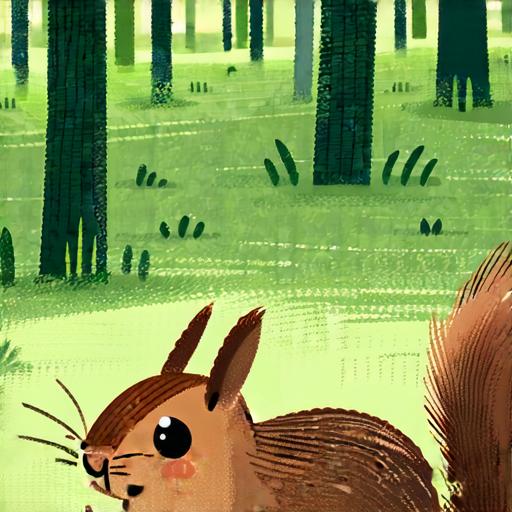} &\includegraphics[width=0.09\linewidth]{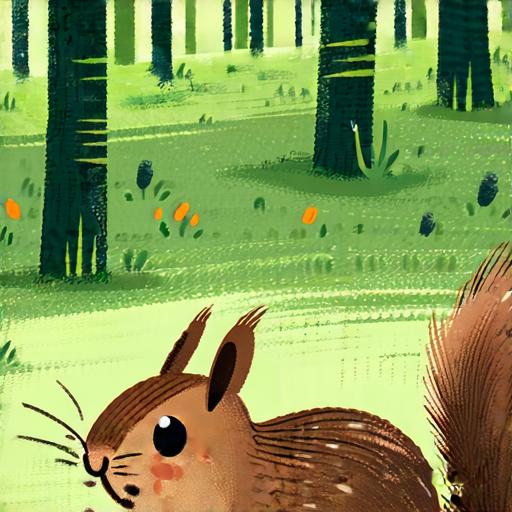} \\
MSE $\epsilon_s$ & 0.00 &0.02 &0.04 &0.06 &0.08 &0.10 &0.12 \\
MSE $z_0$ & 0.00 &0.12 &0.18 &0.22 &0.27 &0.29 &0.31 \\
LPIPS $z_0$ & 0.00 &0.29 &0.39 &0.45 &0.48 &0.50 &0.51 \\
CLIP $z_0$ & 0.00 &0.04 &0.07 &0.05 &0.07 &0.08 &0.08 \\
\raisebox{0.045\linewidth-0.5em}[0em][0em]{Image} & \includegraphics[width=0.09\linewidth]{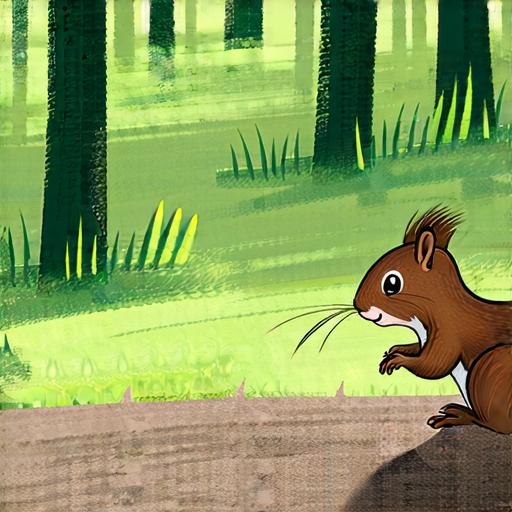} &\includegraphics[width=0.09\linewidth]{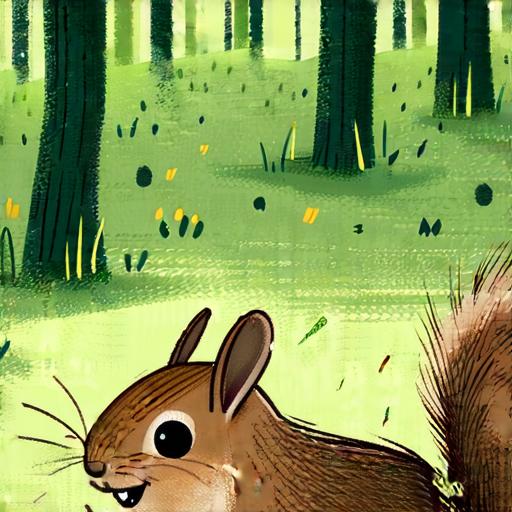} &\includegraphics[width=0.09\linewidth]{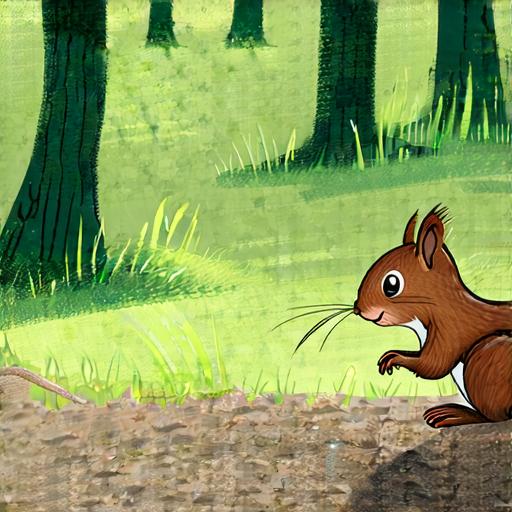} &\includegraphics[width=0.09\linewidth]{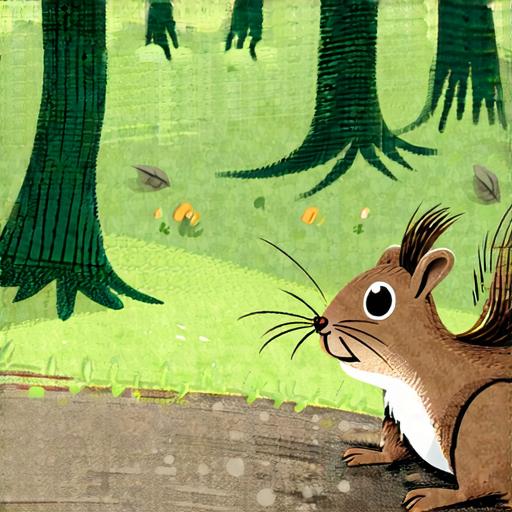} &\includegraphics[width=0.09\linewidth]{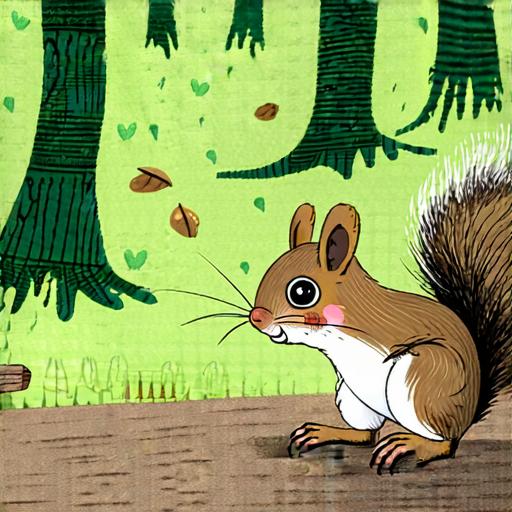} &\includegraphics[width=0.09\linewidth]{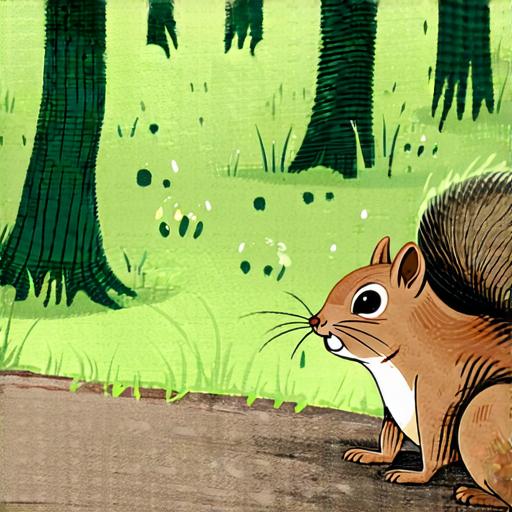} &\includegraphics[width=0.09\linewidth]{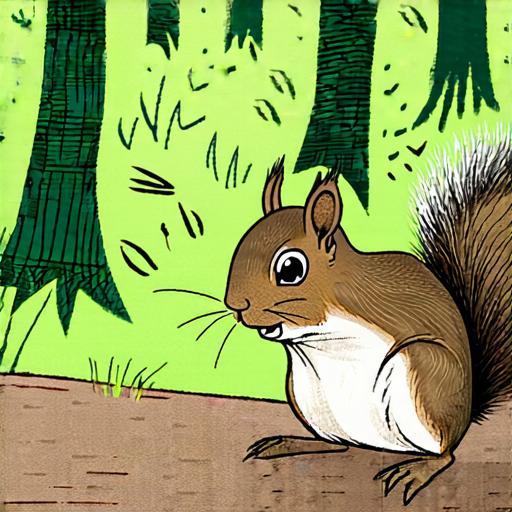} \\
MSE $\epsilon_s$ & 0.15 &0.17 &0.19 &0.21 &0.23 &0.26 &0.28 \\
MSE $z_0$ & 0.33 &0.36 &0.38 &0.40 &0.41 &0.42 &0.43 \\
LPIPS $z_0$ & 0.53 &0.55 &0.56 &0.57 &0.58 &0.58 &0.58 \\
CLIP $z_0$ & 0.10 &0.15 &0.18 &0.21 &0.21 &0.19 &0.18 \\
\raisebox{0.045\linewidth-0.5em}[0em][0em]{Image} & \includegraphics[width=0.09\linewidth]{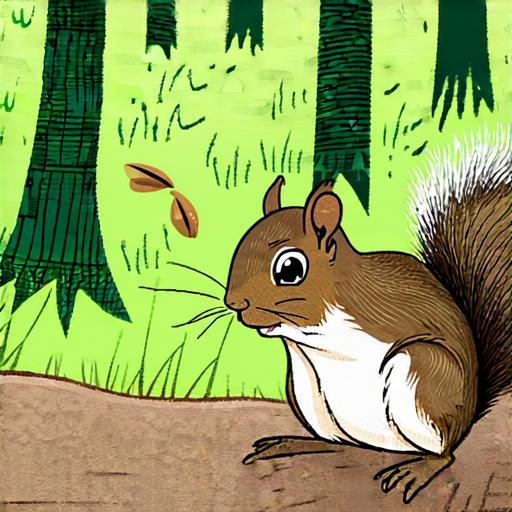} &\includegraphics[width=0.09\linewidth]{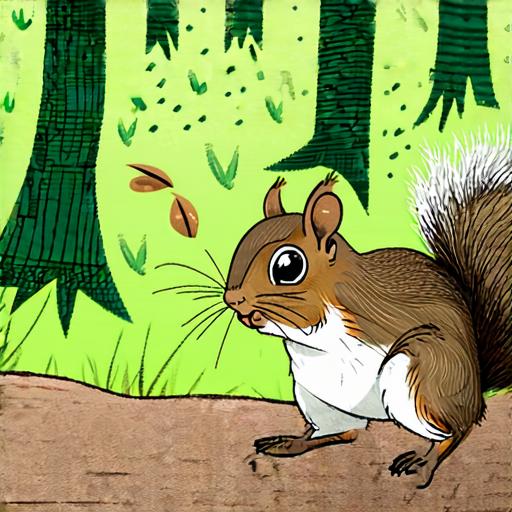} &\includegraphics[width=0.09\linewidth]{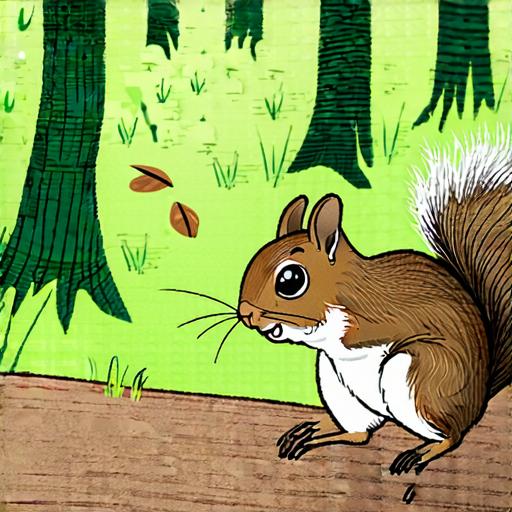} &\includegraphics[width=0.09\linewidth]{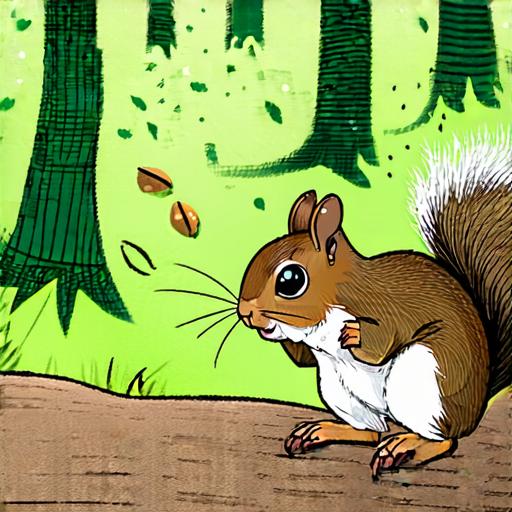} &\includegraphics[width=0.09\linewidth]{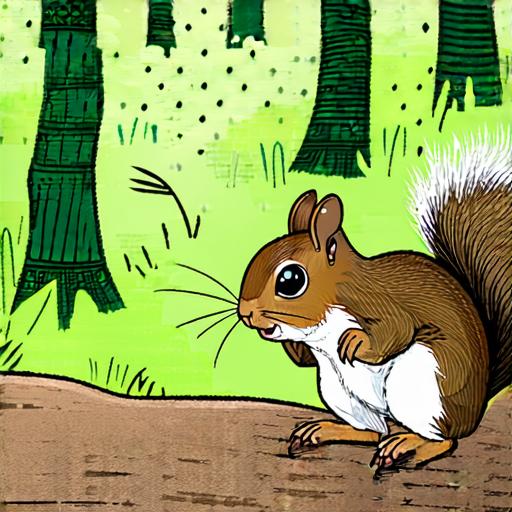} &\includegraphics[width=0.09\linewidth]{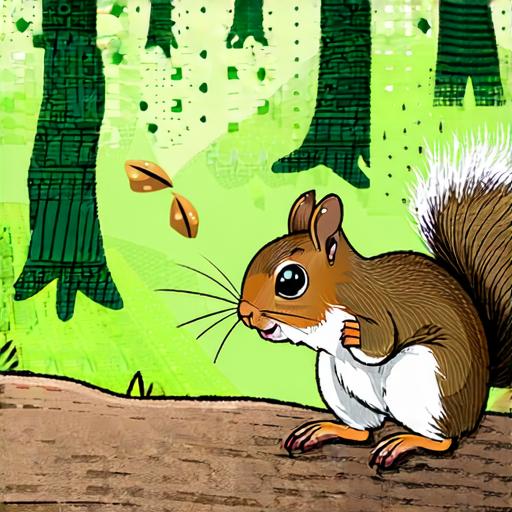} &\includegraphics[width=0.09\linewidth]{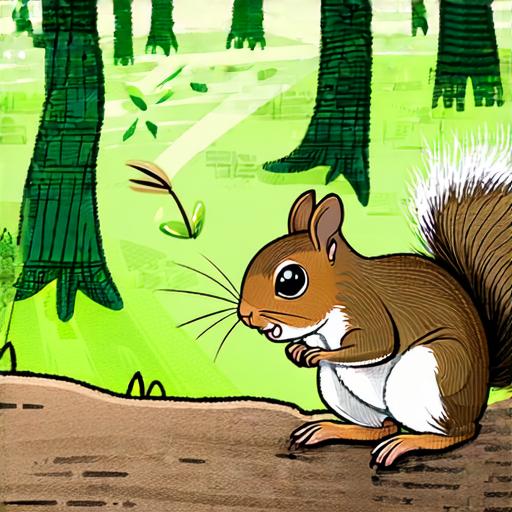} \\
MSE $\epsilon_s$ & 0.30 &0.33 &0.35 &0.38 &0.40 &0.43 &0.45 \\
MSE $z_0$ & 0.44 &0.46 &0.47 &0.47 &0.48 &0.49 &0.49 \\
LPIPS $z_0$ & 0.60 &0.62 &0.62 &0.62 &0.62 &0.62 &0.62 \\
CLIP $z_0$ & 0.18 &0.16 &0.15 &0.16 &0.18 &0.18 &0.19 \\
\raisebox{0.045\linewidth-0.5em}[0em][0em]{Image} & \includegraphics[width=0.09\linewidth]{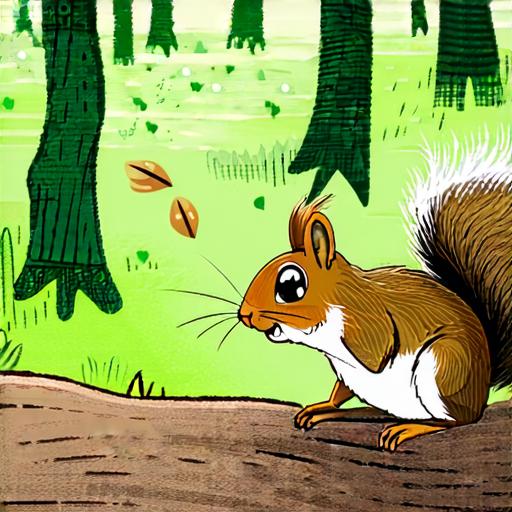} &\includegraphics[width=0.09\linewidth]{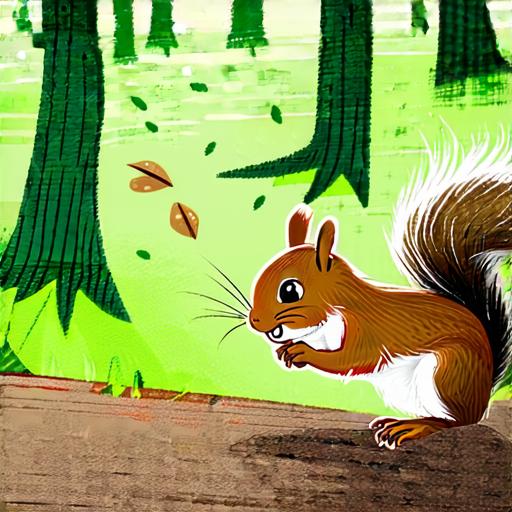} &\includegraphics[width=0.09\linewidth]{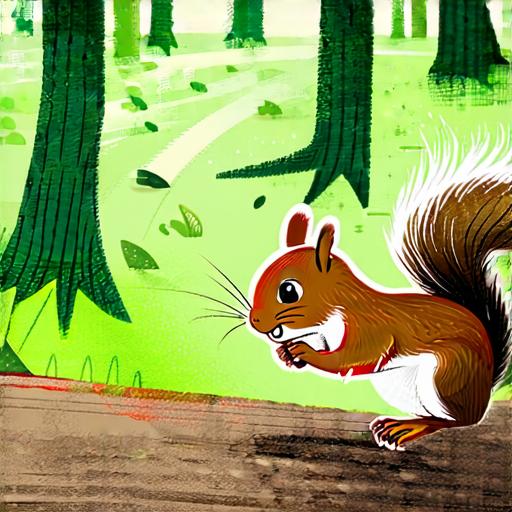} &\includegraphics[width=0.09\linewidth]{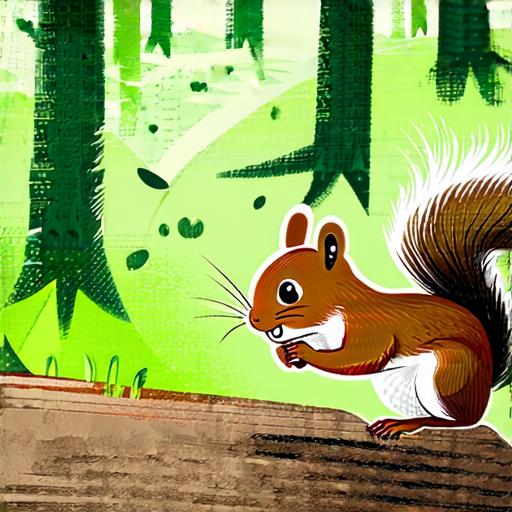} &\includegraphics[width=0.09\linewidth]{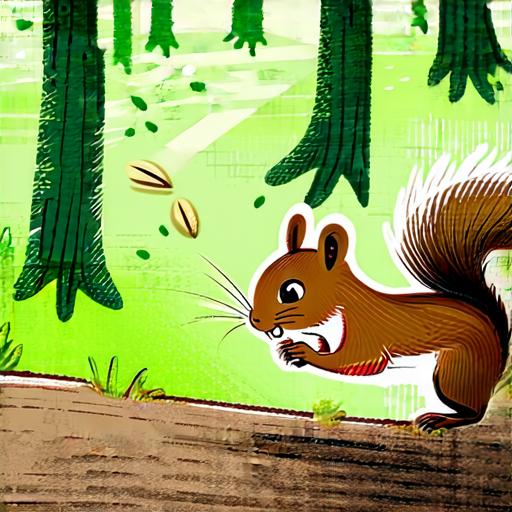} &\includegraphics[width=0.09\linewidth]{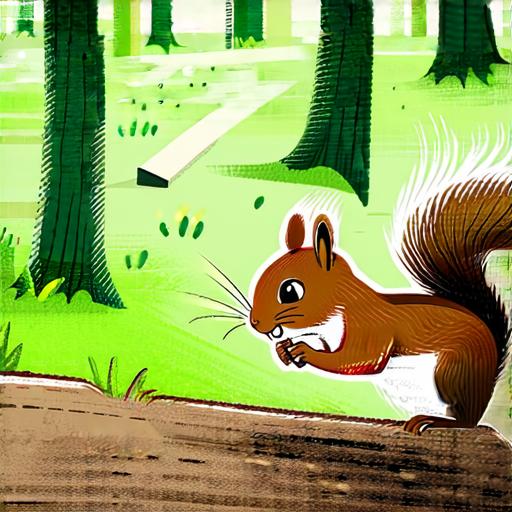} &\includegraphics[width=0.09\linewidth]{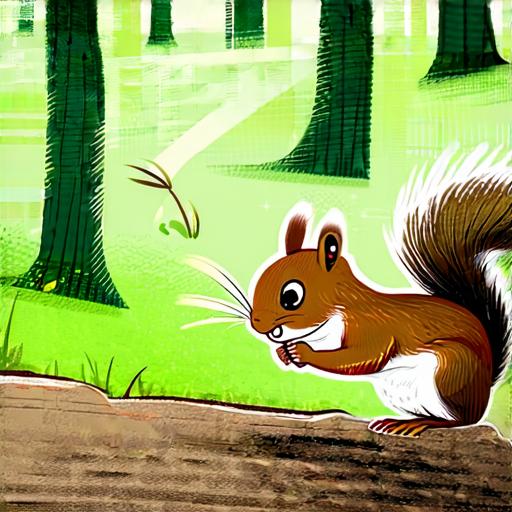} \\
MSE $\epsilon_s$ & 0.48 &0.50 &0.53 &0.56 &0.59 &0.61 &0.64 \\
MSE $z_0$ & 0.50 &0.51 &0.52 &0.52 &0.53 &0.53 &0.53 \\
LPIPS $z_0$ & 0.62 &0.63 &0.63 &0.64 &0.64 &0.64 &0.65 \\
CLIP $z_0$ & 0.20 &0.19 &0.19 &0.19 &0.19 &0.19 &0.21 \\
\raisebox{0.045\linewidth-0.5em}[0em][0em]{Image} & \includegraphics[width=0.09\linewidth]{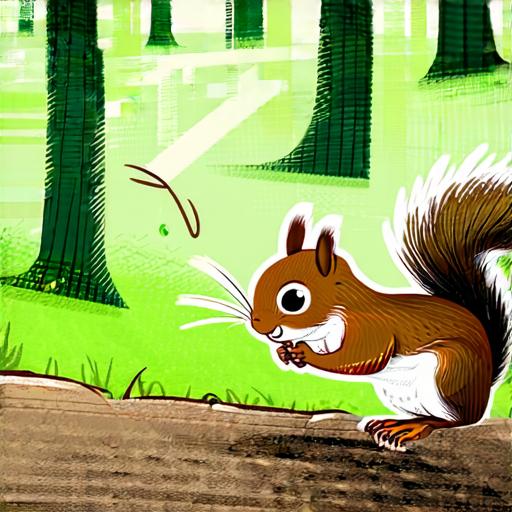} &\includegraphics[width=0.09\linewidth]{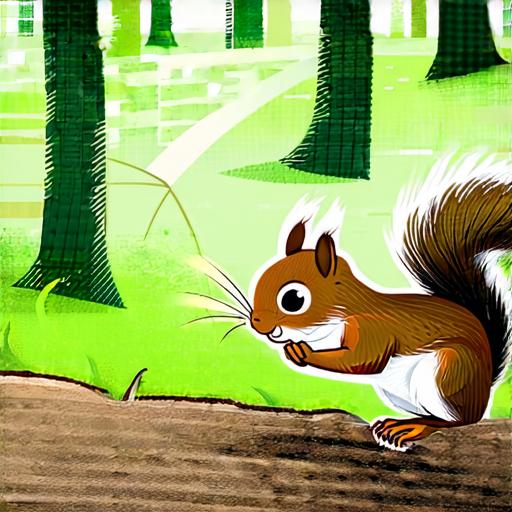} &\includegraphics[width=0.09\linewidth]{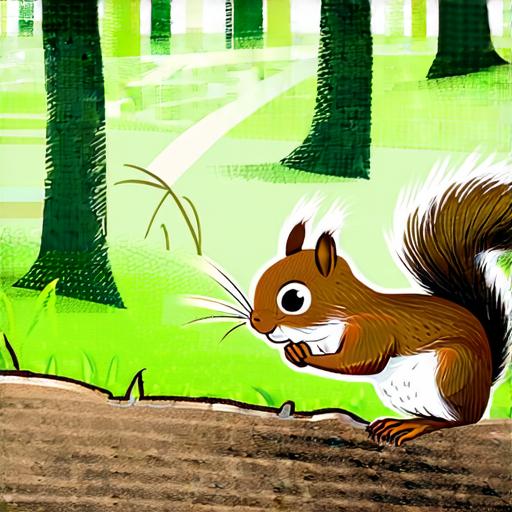} &\includegraphics[width=0.09\linewidth]{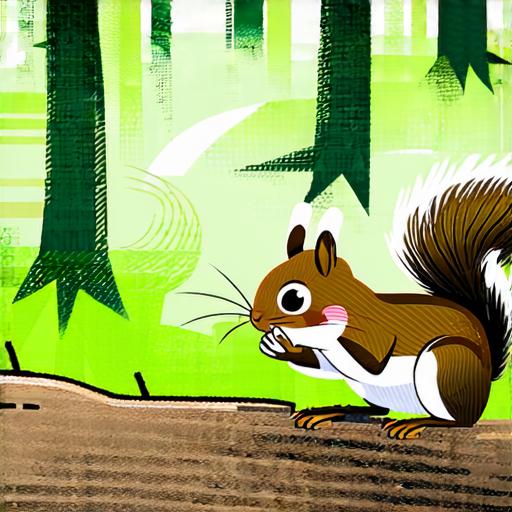} &\includegraphics[width=0.09\linewidth]{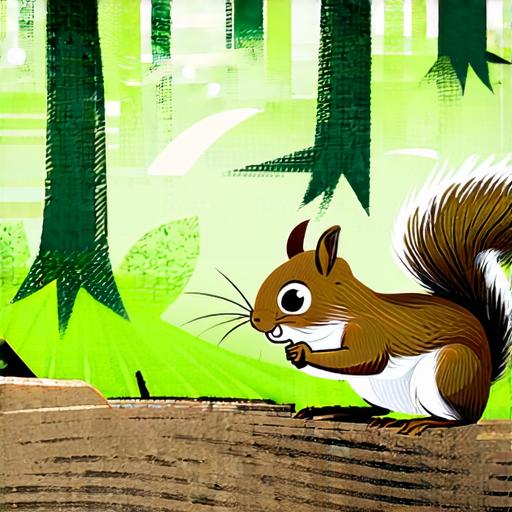} &\includegraphics[width=0.09\linewidth]{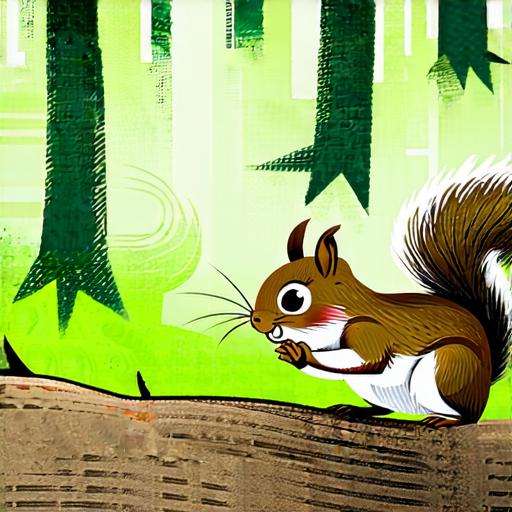} &\includegraphics[width=0.09\linewidth]{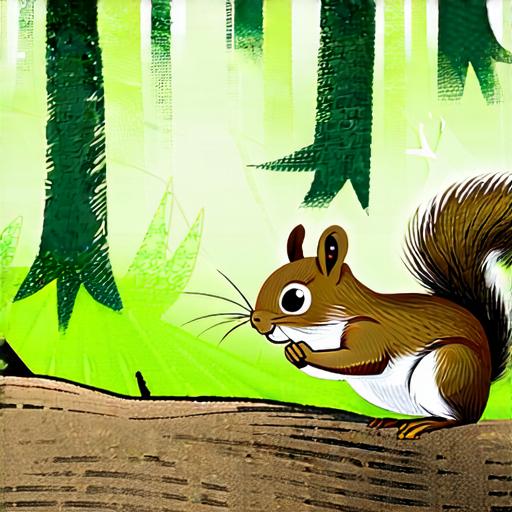} \\
MSE $\epsilon_s$ & 0.67 &0.70 &0.74 &0.77 &0.80 &0.83 &0.87 \\
MSE $z_0$ & 0.54 &0.54 &0.55 &0.55 &0.55 &0.55 &0.55 \\
LPIPS $z_0$ & 0.64 &0.65 &0.65 &0.65 &0.65 &0.66 &0.66 \\
CLIP $z_0$ & 0.21 &0.23 &0.21 &0.24 &0.26 &0.24 &0.24 \\
\raisebox{0.045\linewidth-0.5em}[0em][0em]{Image} & \includegraphics[width=0.09\linewidth]{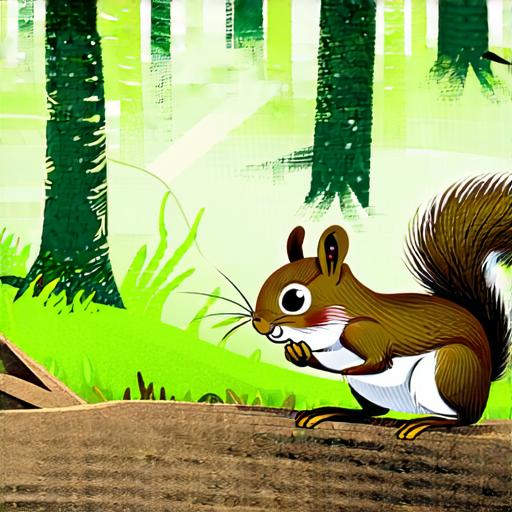} &\includegraphics[width=0.09\linewidth]{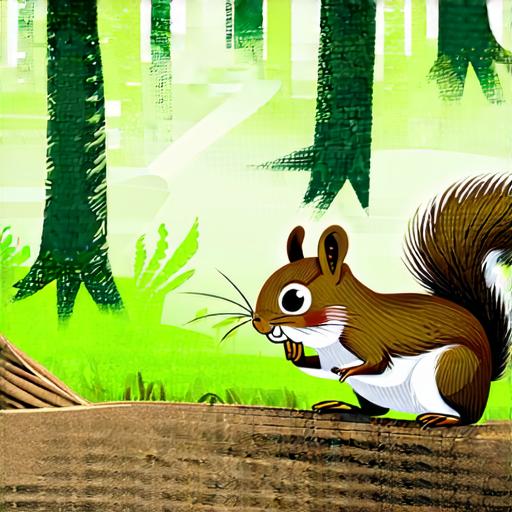} &\includegraphics[width=0.09\linewidth]{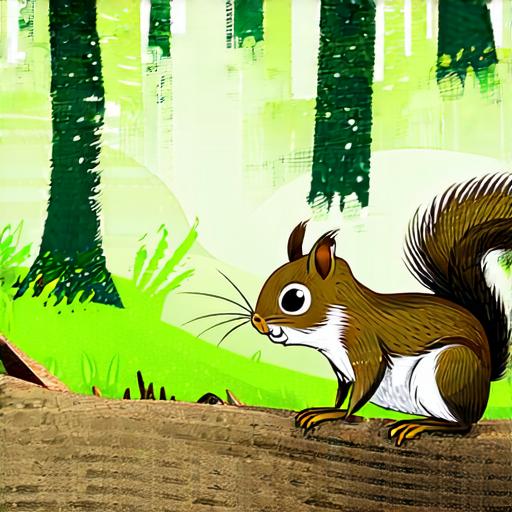} &\includegraphics[width=0.09\linewidth]{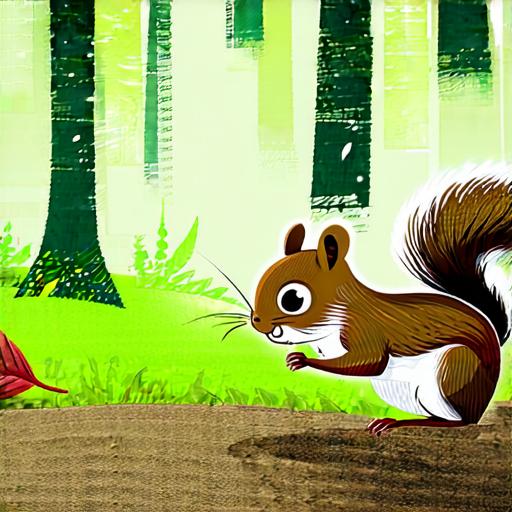} &\includegraphics[width=0.09\linewidth]{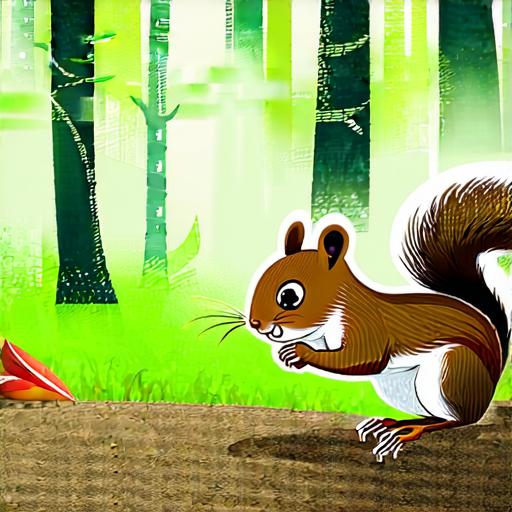} &\includegraphics[width=0.09\linewidth]{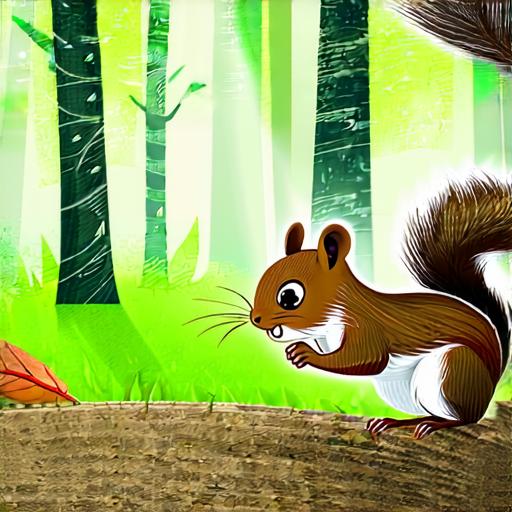} &\includegraphics[width=0.09\linewidth]{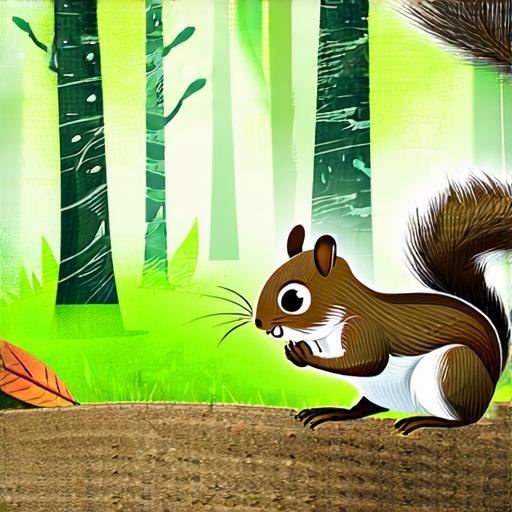} \\
MSE $\epsilon_s$ & 0.90 &0.94 &0.98 &1.02 &1.06 &1.11 &1.15 \\
MSE $z_0$ & 0.55 &0.55 &0.55 &0.55 &0.55 &0.56 &0.57 \\
LPIPS $z_0$ & 0.67 &0.67 &0.68 &0.69 &0.70 &0.71 &0.71 \\
CLIP $z_0$ & 0.22 &0.19 &0.16 &0.15 &0.17 &0.14 &0.15 \\
\raisebox{0.045\linewidth-0.5em}[0em][0em]{Image} & \includegraphics[width=0.09\linewidth]{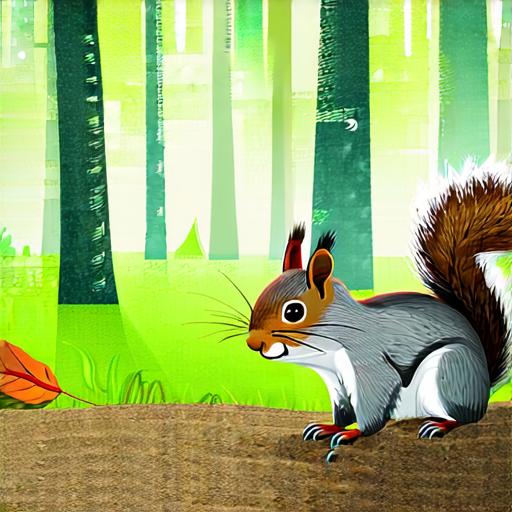} &\includegraphics[width=0.09\linewidth]{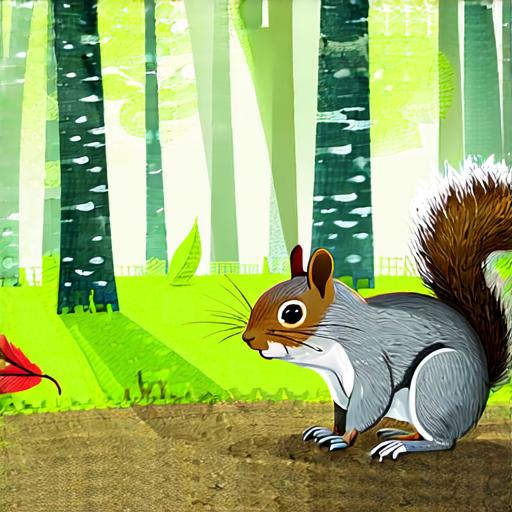} &\includegraphics[width=0.09\linewidth]{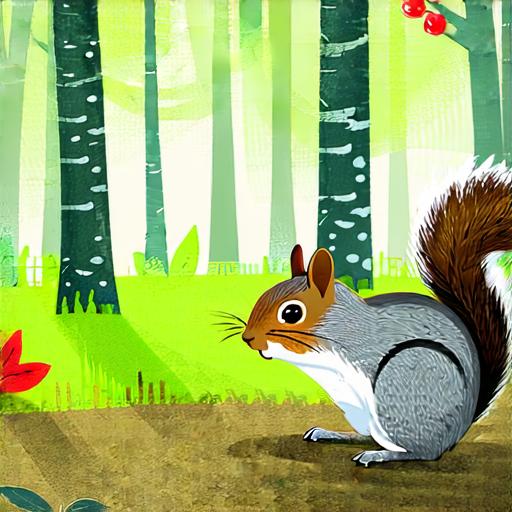} &\includegraphics[width=0.09\linewidth]{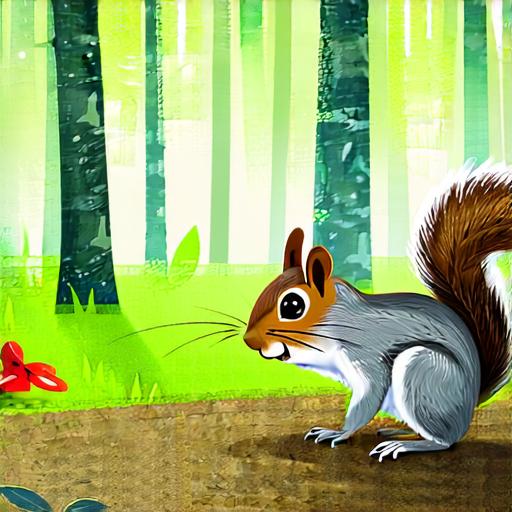} &\includegraphics[width=0.09\linewidth]{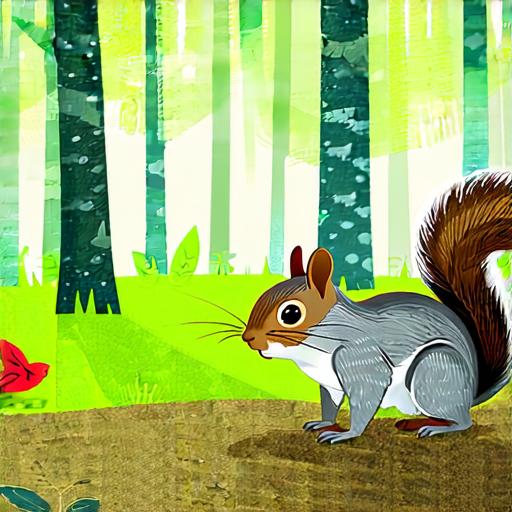} &\includegraphics[width=0.09\linewidth]{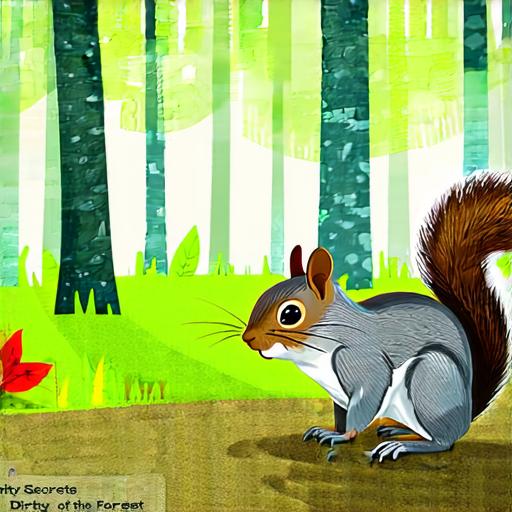} &\includegraphics[width=0.09\linewidth]{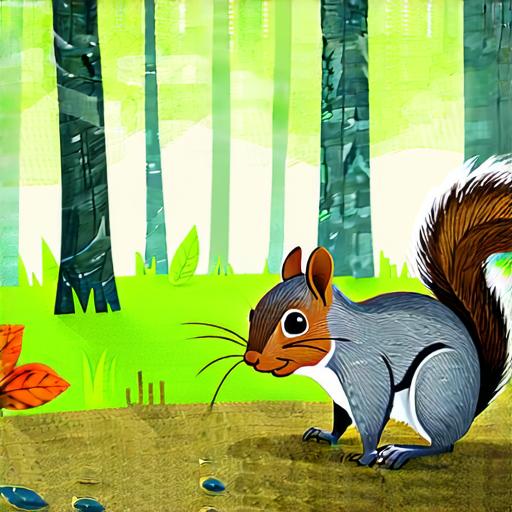} \\
MSE $\epsilon_s$ & 1.20 &1.25 &1.31 &1.37 &1.43 &1.51 &1.60 \\
MSE $z_0$ & 0.59 &0.61 &0.63 &0.63 &0.65 &0.67 &0.68 \\
LPIPS $z_0$ & 0.71 &0.71 &0.72 &0.73 &0.74 &0.75 &0.75 \\
CLIP $z_0$ & 0.19 &0.22 &0.23 &0.24 &0.23 &0.21 &0.20 \\

    \end{tabular}
    \caption{The final image gradually transforms as the initial noise $\epsilon_s$ incrementally changes. All loss functions are calculated with respect to the original image (first image shown). }
    \label{tab:appendix:incremental-noise-change}
\end{table*}

\end{document}